\documentclass[sigconf]{acmart}
\usepackage{booktabs} % For formal tables

\usepackage{url}
\usepackage[draft,inline,nomargin]{fixme}       % During regular working
\usepackage{color}
\usepackage{hyperref}

\usepackage{multirow}
\usepackage{booktabs}
\usepackage{makecell}
\usepackage{tabularx}
\usepackage{threeparttable}
\usepackage{amssymb}

\usepackage{graphicx}
\usepackage{stfloats}
\usepackage{float}
\usepackage{epstopdf}
\usepackage{epsfig}
\usepackage{adjustbox}
\usepackage{tabularx}
\usepackage{subcaption}

\usepackage{tikz}
\usepackage{pgfplots, pgfplotstable}
\pgfplotsset{compat=1.12}
\usepackage{filecontents}
\usetikzlibrary{shapes,arrows,positioning}

\usepackage{xspace}

\newif\ifcomments

\commentsfalse

\ifcomments
\newcommand{\bo}[1]{\textcolor{blue}{ bo: #1}}
\newcommand{\rg}[1]{\textcolor{red}{ rg: #1}}
\newcommand{\ct}[1]{\textcolor{olive}{ ct: #1}}
\newcommand{\dm}[1]{\textcolor{orange}{ dm: #1}}
\else
\newcommand{\bo}[1]{\textcolor{blue}{}}
\newcommand{\rg}[1]{\textcolor{red}{}}
\newcommand{\ct}[1]{\textcolor{olive}{}}
\newcommand{\dm}[1]{\textcolor{orange}{}}
\fi

\newcommand{\post}[2]{\mathtt{post}_{#1}({#2})} %post by user #1 at time #2
\renewcommand{\pm}[3]{\mathtt{msg}_{#1,#2}({#3})} %pm from user #1 to user #2 at time #3
\newcommand{\tpost}{t_p} % time a thread was created
\newcommand{\tpm}{t_m} % time a pm was send
\newcommand{\trstart}{t_l}
\newcommand{\trspan}{d_{l}}
\newcommand{\trend}{\trstart + \trspan}
\newcommand{\tbetween}{\Delta}
\newcommand{\testart}{t_t}%\trstart + \trspan + \trtebetween}
\newcommand{\tespan}{d_{t}}
\newcommand{\teend}{\testart + \tespan}

\newcommand{\model}{\mathcal{M}}

\newcommand{\timessince}{\tau}
\newcommand{\taumax}{\tau_{\text{max}}}
\newcommand{\likelihood}[1]{\mathcal{L}_{#1}}
\newcommand{\binarythreshold}{\Theta}

\newcommand\hmm[1]{\ifnum\ifhmode\spacefactor\else2000\fi>1005 \uppercase{#1}\else#1\fi}

\newcommand{\thread}{\hmm{p}ost\xspace}
\newcommand{\threads}{\hmm{p}osts\xspace}

\newcount{\myi}
\newcommand{\Repeatthis}[2]{%
    \myi=0
    \loop
        \ifnum\myi<#2
        #1
        \advance\myi by 1
    \repeat
}

\begin{document}
\title{Under the Underground: \\Predicting Private Interactions in Underground Forums}

\author{Rebekah Overdorf}
\affiliation{%
  \institution{KU Leuven}
}
\email{rebekah.overdorf@kuleuven.be}

\author{Carmela Troncoso}
\affiliation{%
  \institution{EPFL}
}
\email{carmela.troncoso@epfl.ch}

\author{Damon McCoy}
\affiliation{%
  \institution{NYU}}
\email{mccoy@nyu.edu}

\author{Rachel Greenstadt}
\affiliation{%
  \institution{Drexel University}}
\email{greenie@cs.drexel.edu}

%\author{Anonymized for Submission}

\begin{abstract}
Underground forums where users discuss, buy, and sell illicit services and goods facilitate a better understanding of the economy and organization of cybercriminals. Prior work has shown that in particular private interactions provide a wealth of information about the cybercriminal ecosystem. Yet, those messages are seldom available to analysts, except when there is a leak. To address this problem we propose a supervised machine learning based method able to predict which public \threads will generate private messages, after a partial leak of such messages has occurred. To the best of our knowledge, we are the first to develop a solution to overcome the barrier posed by limited to no information on private activity for underground forum analysis. Additionally, we propose an automate method for labeling posts, significantly reducing the cost of our approach in the presence of real unlabeled data. This method can be tuned to focus on the likelihood of users receiving private messages, or \threads triggering private interactions. We evaluate the performance of our methods using data from three real forum leaks. Our results show that public information can indeed be used to predict private activity, although prediction models do not transfer well between forums. We also find that neither the length of the leak period nor the time between the leak and the prediction have significant impact on our technique's performance, and that NLP features dominate the prediction power. 
%to predict the existence of private interactions from public  to provide analysts with valuable insights into the activities of these forums even when leaks are not available. To this end we frame the private interaction prediction as a machine learning problem in which the leaks inform the training data. We develop a novel method for modeling private interactions in these forums and demonstrate the utility of our model using data from three real forum leaks. We show that our model can be tuned to predict which users are likely to receive messages as well as which particular public posts are likely to spawn private responses. We also show that our model is not only useful at prediction but can also be used to automate the labelling of posts, significantly reducing the cost of our approach in presence of real unlabelled data. 
\end{abstract}

\maketitle

\section{Introduction \label{uf:intro}}

Underground forums facilitate connections between criminals, providing them with a platform to buy, sell, and trade illicit goods and services, such as stolen information, account hacking services, and hacking tools. These forums also enable cybercriminals to discuss illicit topics and share information, providing
%The end result is that underground forums facilitate 
an ecosystem in which criminals can decompose large tasks, such as profiting from spam~\cite{clicktrajectory}, into smaller subtasks that individuals or organizations can focus on solving~\cite{weis15thomas}. Much like in the legitimate business world, this subdividing of responsibilities results in increased efficacy and innovation. % within a particular specialty.  

Understanding structure of these forums is essential to comprehending the cybercriminal pipeline and how these organizations and individuals behave~\cite{weis15thomas}. This has been the focus of significant research~\cite{franklin2007inquiry,weis15thomas,motoyama2011analysis,holt2012examining} as well as several industrial ``threat intelligence'' firms which study cybercriminal forums in order to sell information about emerging attacks and threats. % to various organizations. 
For instance, monitoring of these forums has revealed massive data leaks from Target~\cite{krebs-target2} and Yahoo~\cite{yahooLeak}. Analysis of private parts of the forums have led to making connections which exposed groups of colluding underground forum members~\cite{kerbsDDoS}.

Previous studies of these forums have shown that while public messages are primarily used to advertise goods and services, private messages are a stronger indicator of potential transactions~\cite{motoyama2011analysis}. Thus, studying the private interactions is crucial to fully understand the underlying economy. However, such study has so far been only possible for a few cybercriminal forums for which private messages have been publicly leaked. These leaks are rare and only provide a snapshot of the forum. Thus, they cannot be considered a continuous source of information about private communications.
To address the problem that the absence of private messages leaks poses to forum analysts, in this paper we explore methods to infer the existence of private interactions based on information that is publicly available. We first show that the graph of people communicating privately has little overlap with the graph derived from public interactions. Thus, the public interactions cannot be used to directly infer details about private messages. Instead, we propose a method using supervised learning to predict if a public \thread will result in a private interaction % as well as determine which types of \threads are more likely to cause a private message. Our method 
using the content of the posts and metadata about both the user who created the post and the post itself as features.  

We evaluate our approach on three underground forums for which we have the leaked private message data as well as public \threads for the same time period. Our results show that our method performs significantly better than random. We find that metadata features contain a lot of predictive power individually, however, in aggregate, features related to a \thread's content make these content-based features more influential in classification. Our results show that neither the length of the leak period nor the time between the leak and the prediction have significant impact on our technique's performance, suggesting that activity in the forum is stable over time. Finally, we also find that for the prediction of private interactions to be useful, some training data from the target forum is required. Our methods are useful for analysts to triage a forum and focus additional investigations on likely interesting public \threads.

Determining the ground truth of which private messages are related to a public \thread is not only very costly, but also difficult to conclude even when performed manually. We address this challenge by developing a method of automatically determining the likelihood that a private message is related to a public \thread. For each \thread we then sum these likelihoods and take this sum to label the likelihood that the \thread has an associated private reply during the leak in which we possess both private and public messages. This method is not only useful for reducing the amount of effort required to train our technique on new forums, but also mitigates the difficulty of hand labeling which private messages are replies to public \threads. Our automated labeling has as input a parameter $\binarythreshold$ that tunes the threshold for determining if a \thread is labeled as positive or negative. This inherently affects which features are prioritized: those related to the \thread initiator ($\binarythreshold=0$), or characteristics of the \thread itself (high $\binarythreshold$). Our manual evaluation confirms that high values of $\binarythreshold$ focus the labeler on private messages sent close to the \thread publication time, thus highly likely to be related to it, whereas 
when the $\binarythreshold$ parameter is low, (i.e., it uses any message received by the user), the labeler detects whether the user is likely to receive private messages in general, independent of any particular post.

In summary, our contributions are as follows:

\vspace{1mm}\noindent \checkmark \textbf{We provide the first analysis of the relation between public and private communications in Underground forums.} We show that there is little overlap between the private graph of people communicating in private and the public graph derived from public \threads. Less than 3\% of the users that communicate in public exchange private messages, and roughly 50\% of the users that communicate in private ever commented together on a public \thread. We conclude that there is no straightforward relation between public and private communications in underground forums. 

\vspace{1mm}\noindent\checkmark\textbf{We develop a supervised machine learning based method to predict which public \threads are likely to trigger private messages.} Our experiments show that the AUC of our classifier ranges from 0.65---0.77, depending on the parameters, when training and testing is performed on the same forum. The AUC decreases to 0.60 when testing in another forum. Also, our feature analysis concludes that the content of the post has better predictive power than its metadata, even though the latter still encode a lot of information regarding the likelihood of receiving a response.

\vspace{1mm}\noindent\checkmark\textbf{We model the delay distribution of private messages sent to a user after they create a public \thread.} We observe a spike in the volume of private messages sent to a public \thread creator that degrades at an exponential rate within a few hours of the publishing of the \thread. We model the distribution of this delay as a mixture of exponentials and estimate its parameters, fitting this function to the delays of the incoming messages for each leak on each forum. This distribution allows us to determine the likelihood that a private message is related to a public post. 

\vspace{1mm}\noindent\checkmark\textbf{We develop a method for automatically labeling \threads that are likely to trigger private messages.} This method assigns weights to \threads to express the uncertainty of the relation between \threads and private messages observed in manual labeling. We perform some manual validation by labeling pairs of \threads and private messages sent the the \thread creator. We find that when we look at \threads that were labeled negative by our labeler that 83\% of the sampled privates messages were not related to the post. 

%\item \textbf{We test the developed method across similar forums.} We achieve better than random chance accuracy on the cross-domain analysis when applying the same methods that work well within the same forum. \dm{might want to remove this one unless we have a number that shows it is useful.}

\section{Related Work}

We provide a brief overview of related work on forum analysis and predictive machine learning.

%Prior studies make use of leaked private message data to draw conclusions about how these forums organize and operate. While this type of analysis is critical to understanding these forums, it is not possible without a full leak of a forum, something that is not commonly available. This section takes first steps towards modeling these private messages and understanding how they are triggered. Overall, our aim is to determine if a user who posts a public \thread will receive any private messages as a result.

\subsection{Underground Forum Analysis \label{sec:pmbackground}}

Initial studies of underground forums showed that cybercriminals with specialized skills cooperate and trade different products and services~\cite{thomas2006underground,fallmann2010covertly,franklin2007inquiry,motoyama2011analysis}. Thus, analyses of these forums can be used to understand how these criminals interact with each other and what goods and services are exchanged~\cite{weis15thomas}.
Prior research has aimed to understand why cybercriminals organize by either analyzing private messages~\cite{afroz2014doppelganger,afrozhonor,yip2013forums}, or evaluating self reported studies of members of these communities~\cite{holt2012examining}. Most of this prior research based parts of their analysis or evaluation on private information from forum leaks.
%Text from private messages can be used to link duplicate accounts within the same forum~\cite{afroz2014doppelganger}, which is useful for understanding the full range of activities a person engaged in on a forum. 

Additional research on private messages focuses on how these forums are able to remain operational. Afroz et al.~\cite{afrozhonor} argued that trust management on underground forums is similar to sustaining a common pool resource. However, the success of markets does not indicate the success of individual criminals or vice-versa. Individual criminals have a higher probability of pay-off contingent on their ability to interpret market signals of quality (of both goods/services and individuals selling/buying the same~\cite{decary2013criminals}. As a result, a cybercriminal's ability to succeed or make profits may depend on their location in the network, which is measured by centrality. Individuals with high betweenness centrality have access to more information both quantitatively and in terms of diversity~\cite{decary12}, while individuals with higher degree centrality (for private messages) received more responses for public posts on Carders~\cite{motoyama2011analysis}.  For example, examination of Russian malware writers noted that individuals with higher technical skills were more centrally located~\cite{holt2012examining}. Simultaneously, Dupont  examined a co-offending network of 10 cybercriminals and noted that the more popular criminal did not control the most botnets~\cite{dupont2012skills}. From an enforcement perspective, focusing on degree central criminals is efficient in the former case but not in the latter. Examining the correlation between various centrality measures  on underground forums would illuminate the structural properties of the market and thereby inform deterrence measures~\cite{xu2005crimenet}. Thus, the centrality of a cybercriminal may influence their ability to succeed or make profits. This type of analysis, however, cannot be completed without a leak of the private messages. Our proposed method could potentially enable some of this analysis for longer periods of time based training data from earlier forum leaks. 

\subsection{Predictive Machine Learning}
%The EMBERS system forecasts civil unrest using open source indicators~\cite{embers}.
%\url{https://dl.acm.org/citation.cfm?id=2623373}

Several prior studies that have used machine learning to predict a future event. The EMBERS system forecasts civil unrest using open source indicators~\cite{embers}.
Predicting if a vulnerable website will be compromised and turn malicious based on the if
the vulnerability is being commonly exploited and the profile of the website~\cite{soska14usenix}.
%\url{https://www.usenix.org/system/files/conference/usenixsecurity14/sec14-paper-soska.pdf}.
Predicting if an organization will suffer a data breach based on their externally observable 
security profile and properties of the organization that might make them a more attractive target~\cite{liu15usenix}.
%\url{https://www.usenix.org/system/files/conference/usenixsecurity15/sec15-paper-liu.pdf}.
Our method using similar methods of publicly observable information to infer and predict private 
interactions that are useful for understanding cybercriminal forums.

Machine learning approaches have also been explored, among others, to predict reversions, promotions, and downvotes of user-generated content in web-based communities such as Slashdot~\cite{brennan-wikiai2010} and Wikipedia~\cite{wikisym,collaboratecon}. These approaches have shown that user reputation features and contextual metadata are useful in these systems. The best of our knowledge, our method is the first attempt to tackle this difficult problem of predicting which public \threads will generate private interactions.

\section{Predicting private interactions \label{sec:predicting}}
%In this section we first formalize the problem of predicting which public information spurs private interactions in underground forums. Then, we present a machine learning-based approach to implement such prediction. Additionally, we provide an automated method to efficiently labeling public information that does or does not generate private interactions to use for training.

\subsection{Problem statement}
Let us consider a forum in which $N$ users $u_1,\ldots,u_n$ communicate with each other. A user $u_i$ can perform the following actions: publish a \thread, % at time $\tpost$, denoted as $\post{i}{\tpost}$; 
reply to a \thread, or send a private message to a user $u_j$. % at time $\tpm$, denoted as $\pm{i}{j}{\tpm}$. 
The two former are public, i.e., observable by anyone that has access to the forum, while the latter are private, i.e., only observable to the sender and receiver of the private message. 

While private messages are usually private, they are sometimes made public via so-called \emph{leaks} of the forum database~\cite{forumData}. Such leaks consist of information about the forum users as well as \threads, replies, and private messages during a period of time. We denote the time when the leak starts as $\trstart$, and the duration of the leak period as $\trspan$.

Our goal is to gain understanding on whether the leaked information, i.e., users, \thread, replies, and private messages between $\trstart$ and $\trend$, can be used to predict future private interactions in a forum. More formally, given a \thread $\post{i}{\tpost}$ published by $u_i$ at time $\tpost>\trend$, we aim to predict whether there will be a future private interaction $\pm{*}{i}{\tpm}$, $\tpm>\tpost$, with user $u_i$ as receiver. 

%The previous section, along with existing literature summarized in Section~\ref{sec:pmbackground}, makes use of leaked private message data to draw conclusions about how these forums organize and operate. While this type of analysis is critical to understanding these forums, it is not possible without a full leak of a forum, something that is not commonly available. This section takes first steps towards modeling these private messages and understanding how they are triggered. Overall, our aim is to determine if a user who posts a public \thread will receive any private messages as a result. This relationship between public and private messages, even when all information about them is available, is not clear. We analyze this relationship between public \threads and private messages in Section~\ref{sec:groundtruth}. Section~\ref{sec:pmpredict} follows with the first steps in predicting private message occurrences related to a public \thread. 

\subsection{Approach}

\begin{figure*}[h]
  \centering
    \begin{subfigure}[b]{0.33\textwidth}
        \includegraphics[width=\textwidth]{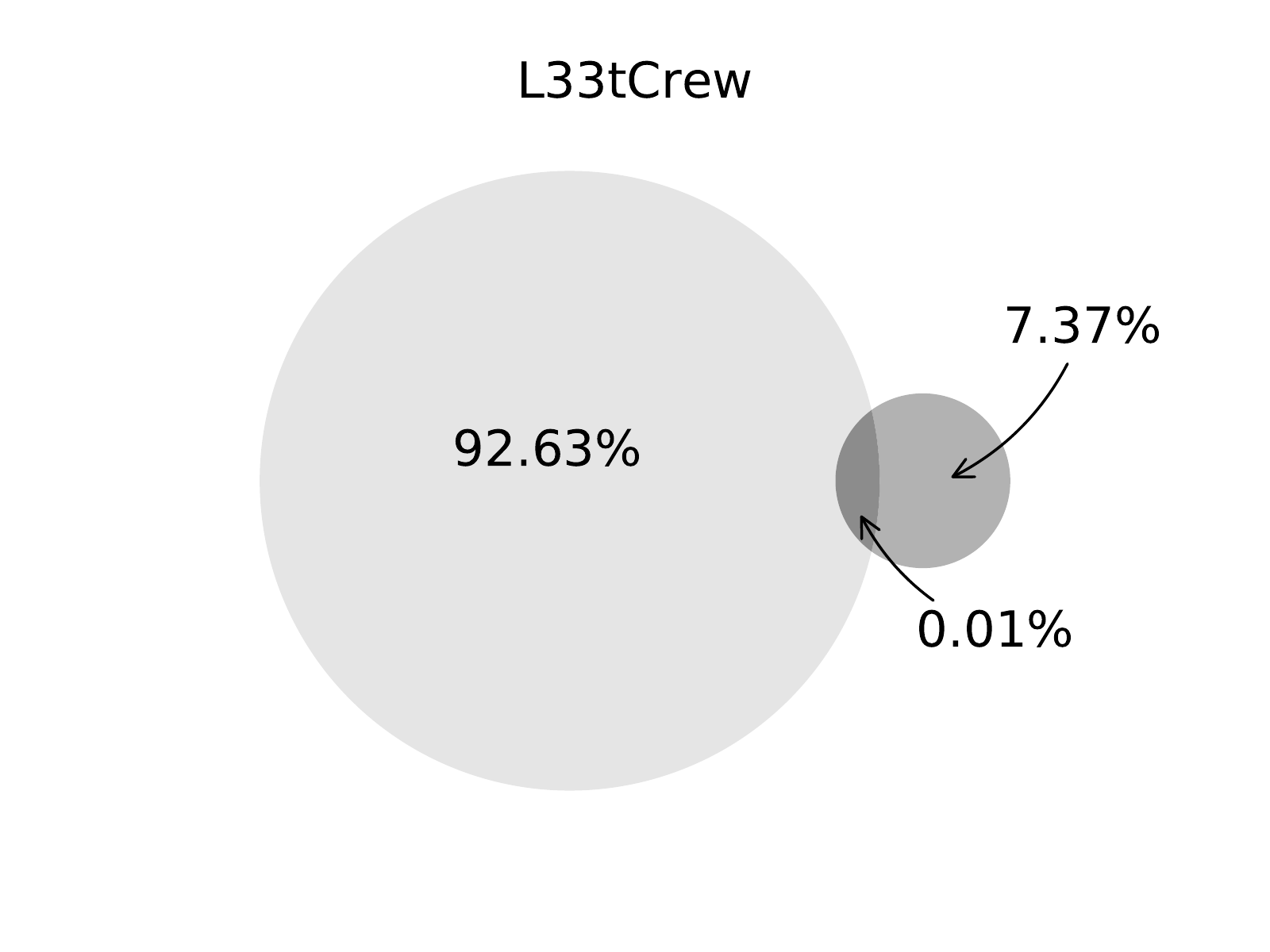}
        %\caption{L33tCrew ($\trspan=?$, $\taumax=?$)}
    \end{subfigure}
    \begin{subfigure}[b]{0.33\textwidth}
        \includegraphics[width=\textwidth]{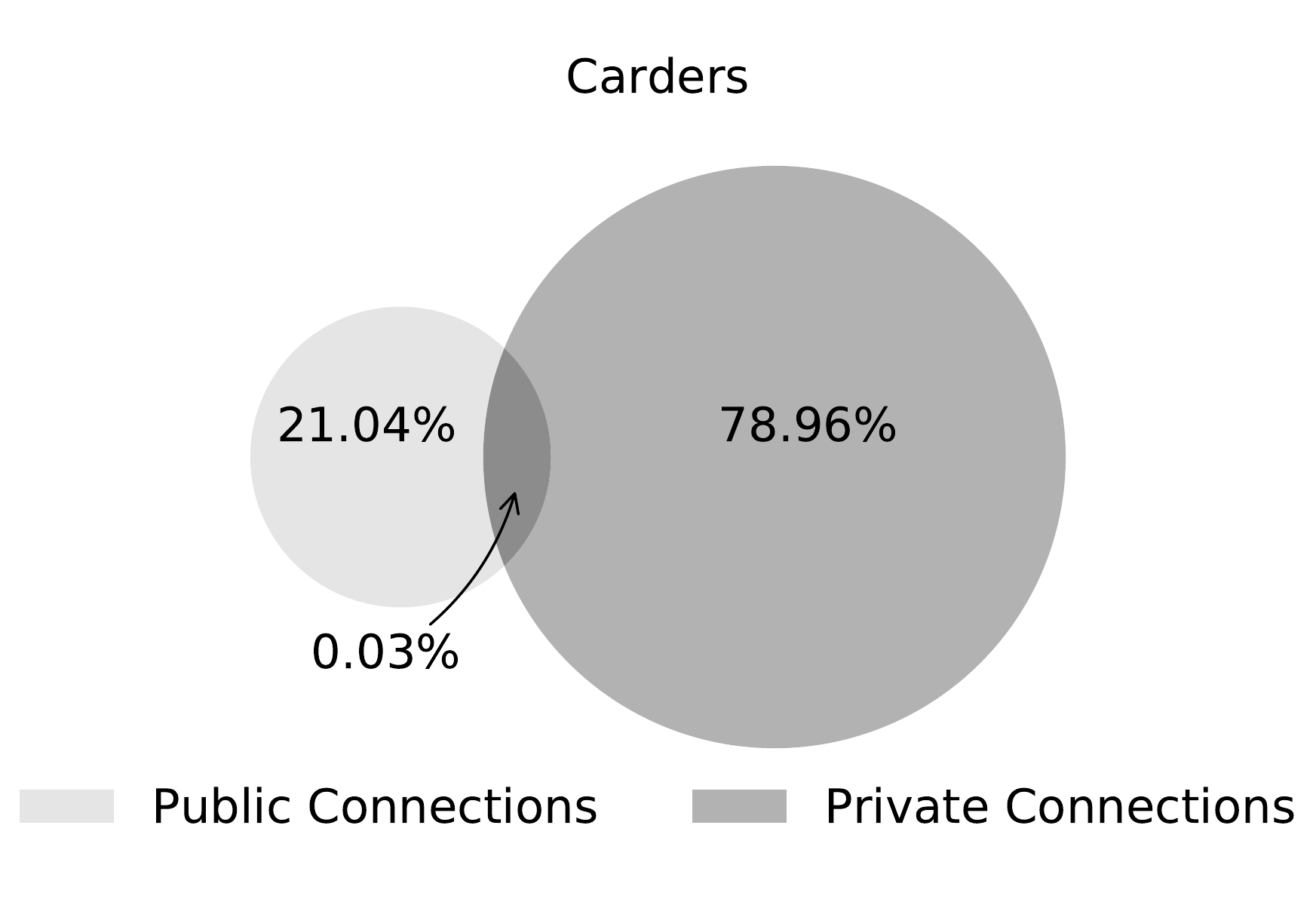}
        %\caption{Nulled ($\trspan=?$, $\taumax=?$)}
    \end{subfigure}
    \begin{subfigure}[b]{0.33\textwidth}
        \includegraphics[width=\textwidth]{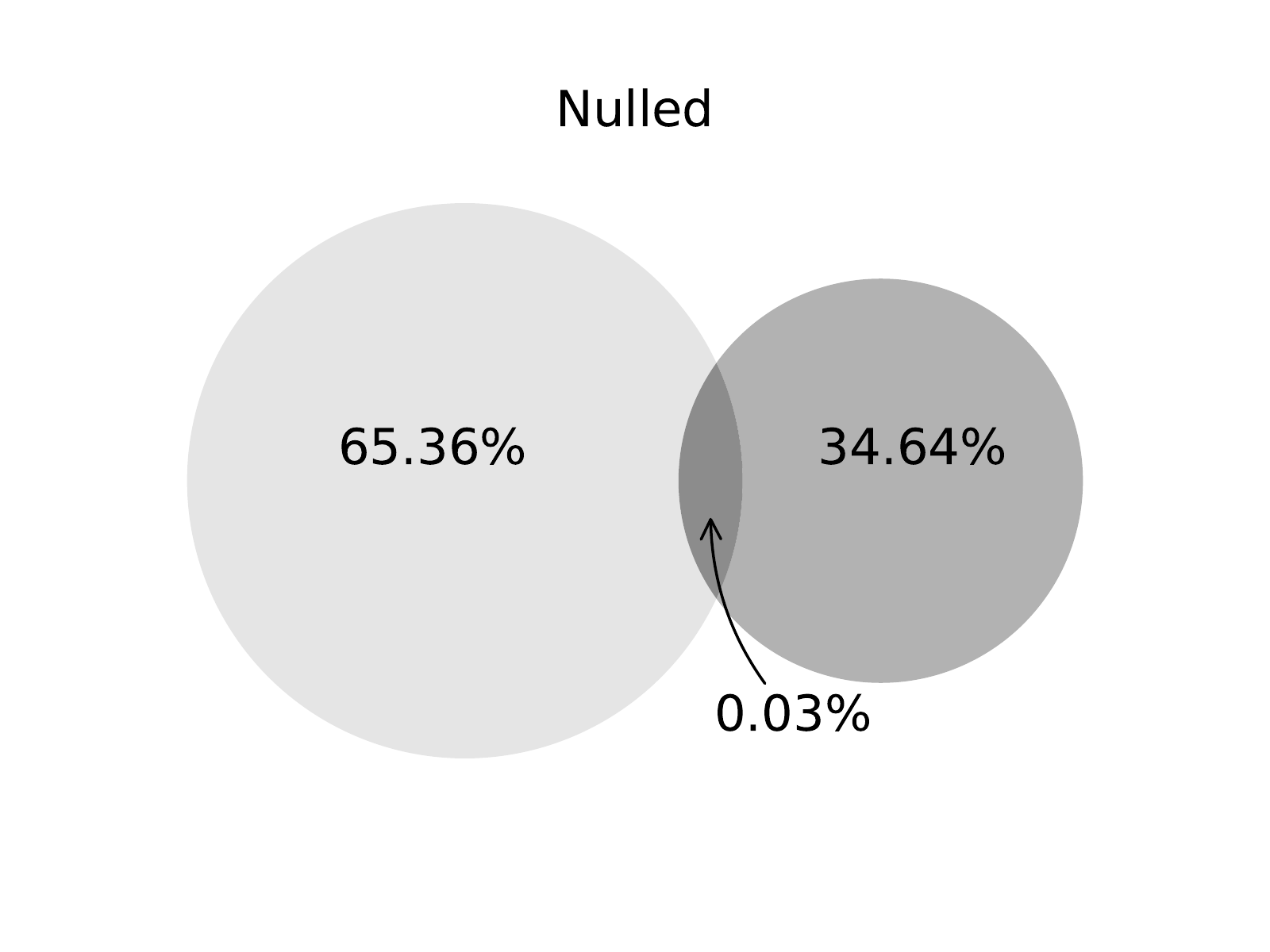}
        %\caption{Carders ($\trspan=?$, $\taumax=?$)}
    \end{subfigure}
   \caption{Similarity of private and public connections. In this analysis a public connection is made from $u_i$ to $u_j$ when a $u_i$ comments in a \thread created by $u_j$, and a private connection is made from $u_i$ to $u_j$ when $u_i$ sends a private message to  $u_j$.\label{fig:venndiagrams}}
\end{figure*}

A straightforward approach to answer the above question would be to use information from the public graph, i.e., who communicates with whom in public, to infer which users communicate in private. However, a preliminary analysis of the available underground forums leaks immediately shows that private and public graphs are far from being related. 

Less than 5\% of users that participate in the same public thread ever communicate in private. Figure~\ref{fig:venndiagrams} shows the overlap of connections made in private versus public for every connection observed during the forum leaks. For all forums, at most 3\% of connections made from one user to another in private also occur in public. 
On Carders, the forum with the \emph{highest} number of private messages proportional to the public interactions, if we only look at the users that do communicate in private, we see that, roughly, only 50\% of them ever commented at the same public thread at some point over the entire length of the leak. 

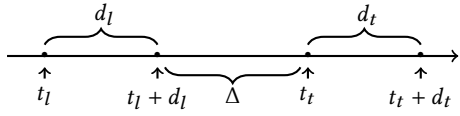
\begin{figure}[h]
\centering
\begin{tikzpicture}[scale=1]

\newcommand{\hei}{0.3}
\newcommand{\heioff}{0.15}

\draw [thick,->] (1.5,0) -- (7.5,0);

%\draw (1,0)node[] {{\textbullet}};
\draw (2,0)node[] {{\textbullet}};
\draw (3.5,0)node[] {{\textbullet}};
\draw (5.5,0)node[] {{\textbullet}};
\draw (7,0)node[] {{\textbullet}};
%\draw (12,0)node[] {{\textbullet}};

%\node[align=center,anchor=south] at (1,\hei) {Forum Start};
%\draw [thick,<-] (1,\heioff) -- (1,\hei);

\node[align=center,anchor=north] at (2,-\hei) {$\trstart$};
\draw [thick,<-] (2,-\heioff) -- (2,-\hei);

\node[align=center,anchor=south] at (2.8,\hei) {$\trspan$};
\draw [thick,decorate,decoration={brace,amplitude=6pt,raise=0pt}] (2,\heioff) -- (3.5,\heioff);

\node[align=center,anchor=north] at (3.5,-\hei) {$\trend$};
\draw [thick,<-] (3.5,-\heioff) -- (3.5,-\hei);

\node[align=center,anchor=north] at (4.5,-\hei) {$\tbetween$};
\draw [thick,decorate,decoration={brace,amplitude=6pt,raise=0pt,mirror}] (3.6,-\heioff) -- (5.4,-\heioff);

\node[align=center,anchor=north] at (5.5,-\hei) {$\testart$};
\draw [thick,<-] (5.5,-\heioff) -- (5.5,-\hei);

\node[align=center, anchor=south] at (6.3,\hei) {$\tespan$};
\draw [thick,decorate,decoration={brace,amplitude=6pt,raise=0pt}] (5.5,\heioff) -- (7,\heioff);

\node[align=center,anchor=north] at (7,-\hei) {$\teend$};
\draw [thick,<-] (7,-\heioff) -- (7,-\hei);

%\node[align=center,anchor=south] at (12,\hei) {Forum End};
%\draw [thick,<-] (12,\heioff) -- (12,\hei);

\end{tikzpicture}

\caption{\emph{Leak} period ($\trstart$, $\trend$) of duration $\trspan$ and \emph{target} period ($\testart$, $\teend$) of duration $\tespan$ separated by a time $\tbetween$. \label{forumtimeline}}
\end{figure}

These results suggest that the relation between public and private interactions is much more complex. To address this complexity, we frame the question of inferring private communication as a supervised machine learning problem. Assume there is a leak with both public and private interactions over a \emph{leak period} between $\trstart$ and $\trend$ of duration $\trspan$, and that the goal is to predict private interactions during a \emph{target period}  $(\testart,\teend)$ of duration $\tespan$, strictly distinct from the leak period, for which we only have access to public information. Figure \ref{forumtimeline} illustrates this timeline. The parameter $\tbetween$ denotes the time passed since the end of the leak period and the start of the target period.
%\ct{think about we really want $\tespan$, anyway say no need to be the same as $\trspan$}

The public and private data available from the leak period can be used as a \emph{training set} to build a model $\model$ that represents the likelihood that a user receives a private message after publishing a \thread. This model uses as features public information about a \thread $\post{i}{\tpost}$ (e.g., characteristics of the sender $u_i$, number of to the \thread replies, \thread content) published in the target period. The prediction performance greatly depends on the features used to train the model. We provide examples of feature sets, and evaluate their performance, in Section~\ref{sec:eval}.

\subsection{Automating labeling \label{sec:groundtruth}}

Note that the available leaks do not contain explicit information about public-private messages correspondence. Thus, one challenge in this problem is that there is no ground truth linking private messages to public threads. Even with all of the leaked information available it is unclear which private messages are related to which public posts or even if a public \thread received any replies. 

One option to solve this problem would be to manually label the data.
This would mean going through each of the \threads published in the leak period and searching through the private messages for related conversations. This process is not only prohibitively expensive, but also difficult. Manual linking must be evaluated in terms of content and in many cases, even for a human, the relationship between public \threads is not always clear, hindering the labeling task. As an example, we did manually label a small set of 330 \threads and message pairs with the goal of determining if they were related. We ended up labeling almost half, 47\%, as \emph{unclear}. Section~\ref{sec:eval} further details this analysis. 

Thus, we develop a method to automatically label \threads in the training set. Acknowledging that the relation between \threads and private messages is fuzzy, and that as a result labels will be noisy, our method does not use binary yes/no labels but assigns a weight to each \thread. This weight effectively models the \thread likelihood of having triggered a private message.
%\ct{@Bekah, please check my text above! ignorant writing about machine learning. I am sure there is much better terminology to be used} \rg{refined slightly}

\begin{figure}[h]
\centering
\begin{tikzpicture}[scale=1]
\newcommand{\hei}{0.6}
\newcommand{\heioff}{0.15}

\draw [thick,->] (1.5,0) -- (9.5,0);

%\draw (1,0)node[] {{\textbullet}};
\draw (2,0)node[] {{\textbullet}};
\draw (3.5,0)node[] {{\textbullet}};
\draw (6,0)node[] {{\textbullet}};
\draw (8.5,0)node[] {{\textbullet}};
%\draw (12,0)node[] {{\textbullet}};

%\node[align=center,anchor=south] at (1,\hei) {$\trstart$};
%\draw [thick,<-] (1,\heioff) -- (1,\hei);

\node[align=center,anchor=north] at (2,-2*\hei) {$\post{i}{\tpost}$};
\draw [thick,<-] (2,-2*\heioff) -- (2,-2*\hei);

\node[align=center,anchor=center] at (2.7,2*\hei) {$\tau_1$};
\draw [thick,decorate,decoration={brace,amplitude=6pt,raise=0pt}] (2.1,5*\heioff) -- (3.4,5*\heioff);

\node[align=center,anchor=north] at (3.5,-\hei) {$\pm{i}{*}{t_{m,1}}$};
\draw [thick,<-] (3.5,-\heioff) -- (3.5,-\hei);

\node[align=center] at (4,1.5*\hei) {$\tau_2$};
\draw [thick,decorate,decoration={brace,amplitude=6pt,raise=0pt}] (2.1,3*\heioff) -- (5.9,3*\heioff);

\node[align=center,anchor=north] at (6,-\hei) {$\pm{i}{*}{t_{m,2}}$};
\draw [thick,<-] (6,-\heioff) -- (6,-\hei);

\node[align=center] at (7,\hei) {$\tau_3$};
\draw [thick,decorate,decoration={brace,amplitude=6pt,raise=0pt,aspect=0.775}] (2.1,\heioff) -- (8.4,\heioff);

\node[align=center,anchor=north] at (8.5,-\hei) {$\pm{i}{*}{t_{m,3}}$};
\draw [thick,<-] (8.5,-\heioff) -- (8.5,-\hei);

%\node[align=center,anchor=south] at (12,\hei) {$\trend$};
%\draw [thick,<-] (12,\heioff) -- (12,\hei);

\end{tikzpicture}

\caption{Timing relationship between posting, $\tpost$, and private messages received by the post creator at $t_{m,1}$, $t_{m,2}$, and $t_{m,3}$. \label{pmtimeline}}
\end{figure}
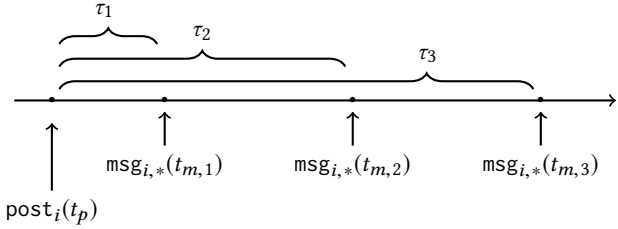

As a first step, we aim to model the relationship between the time when a user $u_i$ publishes a public \thread, $\post{i}{\tpost}$, and when she receives a private message $\pm{*}{i}{\tpm}$. We denote this time difference as $\tau_x$ and compute it as $\tau_x=t_{m,x}-\tpost$, where $x$ indicates the index of subsequent messages received by the \thread initiator in increasing order (see Figure \ref{pmtimeline}). 

\begin{figure}[h]
 \centering
    \includegraphics[width=1\linewidth]{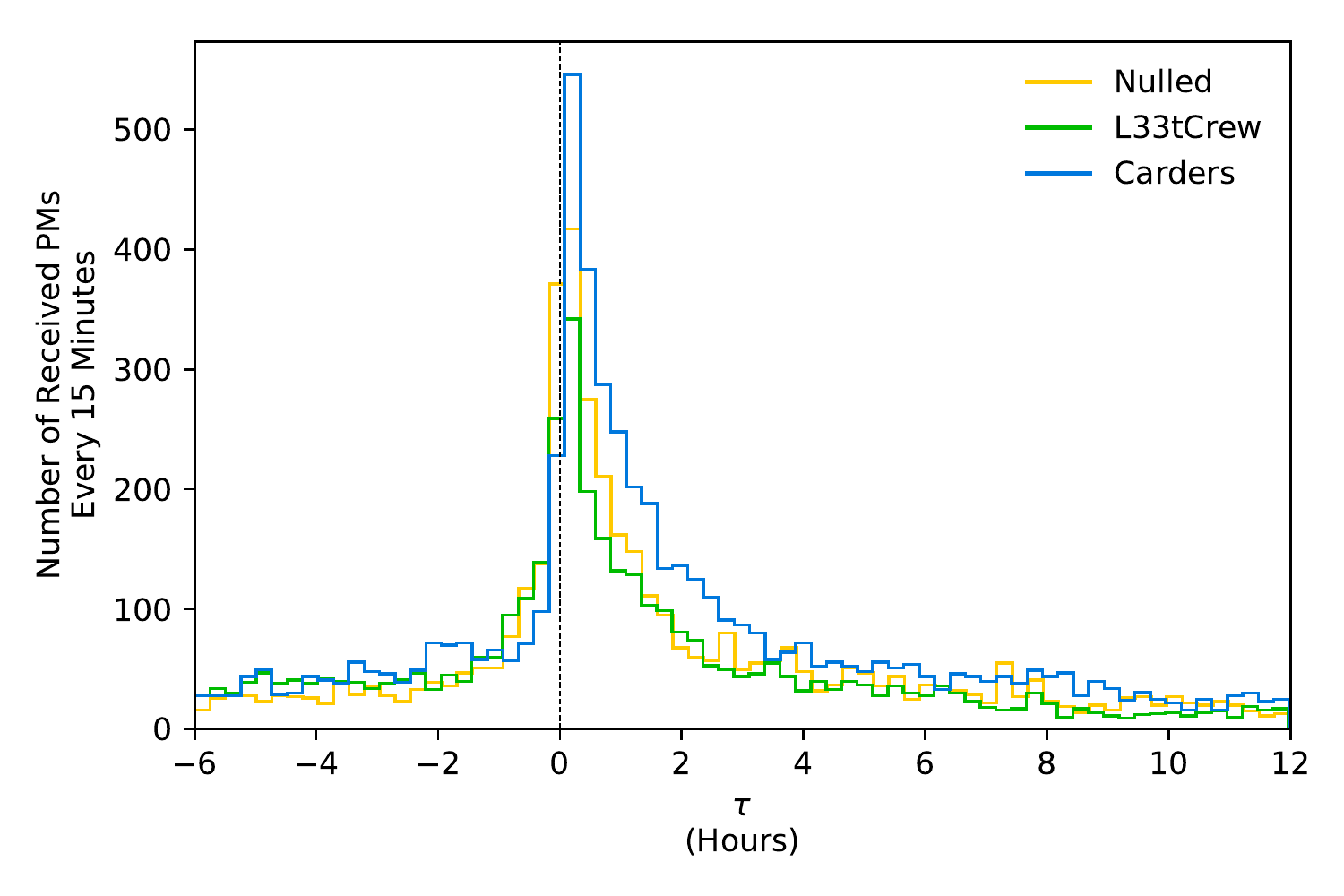}
    \caption{Distribution of intervals $\tau$ between post publication and reception of private messages for each forum over a leak of size six weeks.\label{fig:ts_hist}}% \ct{change axis label. Remove title.} \bo{the axis labels aren't wrong? Fixed legend and title.} }

\end{figure}

To this end we compute the distribution of intervals $\tau$ for all \threads and private mesages in the leak period. The results are shown in Figure~\ref{fig:ts_hist} for the three forums we use in our experiments. %, see Section~\ref{sec:forums}. 
In this figure each bar represents the volume of incoming messages in intervals of 15 minutes across the entirety of a leak period of 6 weeks. We observe a large spike shortly after a \thread is published, which we conjecture is generated by a large amount of messages triggered by such post. Note that users also receive messages when they have not recently posted, as indicated by the long tails of the distribution. Negative values of $\tau$ are a result of private messages received prior to a given \thread publication that can be either responses to previous \threads or spontaneous messages between users. We recall that in this paper we are only interested in predicting messages sent \emph{after} a \thread and therefore in the reminder of this document we disregard negative values of $\tau$.

%, . This spike can be seen on all of the forums in Figure~\ref{fig:ts_hist}.  This figure displays $\timesinceset$ as a histogram where each bar represents the frequency of incoming messages every 15 minutes across the entirety of each of the forums. That is, a histogram of the differences between when a \thread is posted and when the poster of the \thread receives a private message. This shows the influx of messages to users that post public\threads after they do. 

To represent both behaviors, the spike after publication and the low posterior influx, we model the likelihood of a given arrival time of incoming private messages to a user after a \thread as a mixture of two exponentials:
\begin{equation}
f(x) = a_{1}e^{-b_{1}x} + a_{2}e^{-b_{2}x} \,% + c_\timessince
\label{eq:f}
\end{equation}
The first exponential, defined by coefficients $a_1$ and $b_1$, aims at capturing the initial steep decline, while the one defined by $a_2$ and $b_2$ captures the slow decay in the number of messages over time.

The value of these coefficients can be estimated by interpolating the function $f(x)$ over the data points from the training data obtained in the leak period. Note, however, that Figure \ref{fig:ts_hist} also shows the existence of a continuous flow of messages independent of the existence of the post (distribution tail). We model this post-independent message flow as a constant $c$. Then, instead of directly interpolating $f(x)$, we interpolate $f(x)+c$ from which can easily obtain $f(x)$ by substracting the constant. We measure how well the interpolation fits to the data by computing the coefficient of determination $R^2 = 1 - (\sum_i{(y_i - \bar{y})^2} / (\sum_i{(y_i - f_i(x))^2})$ where each $y_i$ is the volume of incoming messages at time $i$. 

The values of the coefficients obtained via interpolation depend on three parameters. Naturally, on the duration of the leak period, $\trspan$, that determines the number of posts and private messages available to construct the histogram that we use as input to interpolation. Secondly, they depend on the maximum time $\taumax$ after a \thread for which a private message is considered to be possibly related to this post. This not only affects the number of private messages available, but also the weight of the distribution tail on the interpolation problem. To avoid setting this value arbitrarily, we infer it empirically by testing values for $\taumax$ and choosing the one for which the values of the coefficients do not change more than $10e-3$ when it is increased. We stress that, though we cap $\tau$ for the interpolation, $f(x)$ is not truncated so as to indicate that \emph{any} message in the future could be related to a \thread, though the likelihood is very small.

Finally, the values of the coefficients depend on the granularity used to compute the histogram representing the distribution. In Figure~\ref{fig:ts_hist} time is arbitraritly divided in 15-minute bins for $\timessince$. For our experiments, we explored a few methods for determining the optimal granularity of the distribution. The \emph{na\"ive} method is to choose, as in Figure~\ref{fig:ts_hist} fix the bin size for any length of the leak period $\trspan$. This method is harmful for smaller leaks, since the bins are too small and many end up empty providing a poor input to the interpolation. Instead, we developed a \emph{balanced} method that determines the bin granularity based on the average number of items desired in each bucket. This method ensures that one has enough quality data points for interpolation, i.e., all bins have a significant number of points, regardless of the duration leak and the level of activity in the forum. 

\begin{figure}[t]
  \centering
        \includegraphics[width=\columnwidth]{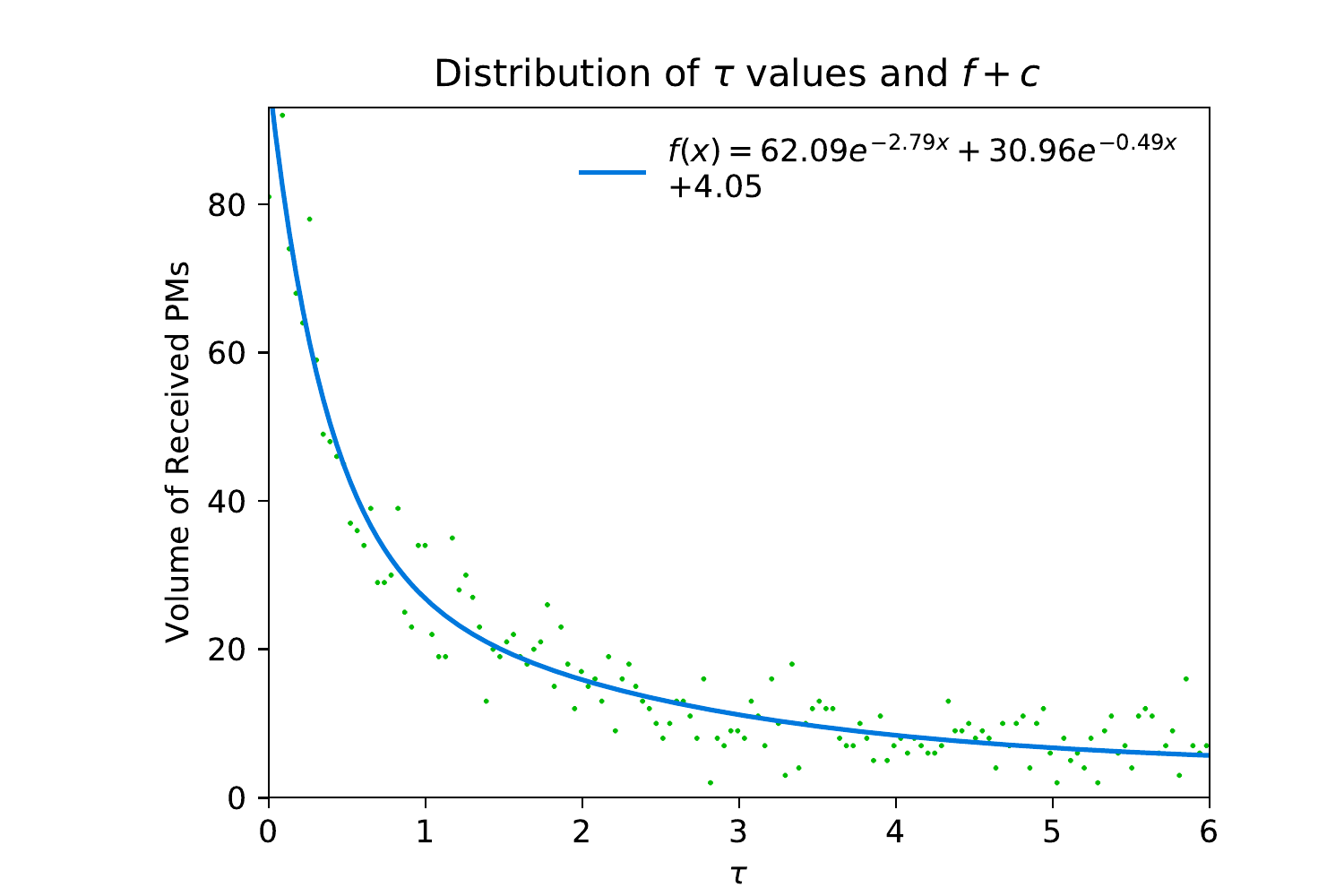}
        \caption{Distribution of $\timessince$ values and $f(x)+c$ for L33tCrew. ($\trspan=7$ weeks, $\taumax=15$ hours, $bucket\_average=4$, $R^2=0.939561$)\label{fig:fexample}}
\end{figure}

We tested both of these methods with a number of inputs, and compared the obtained $R^2$ values. As expected, the smaller leaks saw the greatest improvement with the balanced method, with a $\trspan$ of one week improving from $R^2=0.50$ for one minute bins, and $R^2=0.83$ for five minutes bins, to $R^2=0.96$ with the balanced method with an average of five items per bucket.\bo{I'd \emph{love} a table here if we have space!} 
We show in Figure~\ref{fig:fexample} the result of the interpolation for L33tCrew. 

Since the function $f(x)$ models the likelihood of a private message being sent to a \thread creator after the thread has been created, it can be used to infer a label that represents the core question in this paper: will a user who posts a public \thread receive a private message? For this purpose, we define the aggregated likelihood that a given \thread $\post{i}{\tpost}$ sent by user $u_i$ receives a response as:

\begin{equation}\label{eq:likelihood}
\likelihood{\post{i}{\tpost}} = \sum{f(\timessince_x)}, \,  \forall \timessince_x=t_{m,x} - \tpost, 
\end{equation}
such that $t_{m,x}>\tpost$ and there exists a private message $\pm{*}{i}{t_{m,x}}$ to $u_i$ at that time.

The above definition of the likelihood closely resembles that of the joint probability. However, we note that $f(x)$ is not a probability distribution, thus nor is $\likelihood{\post{i}{\tpost}}$. First, $f(x)$ is fitted to the volume of $\tau$ observed in the training dataset, thus it is possible that the area under the curve does not add up to 1. Normalizing after the interpolation to transform $f(x)$ into a distribution is also not possible given that the support for the exponential goes to infinity.

Once the aggregated likelihood is computed, we assign a binary label to posts using a threshold $\binarythreshold \in \{0,\ldots,\infty\}$ to decide whether the \thread has generated messages, or not. When $\binarythreshold = 0$, i.e., when we evaluate that $\likelihood{\post{i}{\tpost}}>0$, \threads initiated by a user who \emph{ever} receives a message during the leak period are labeled positively. When $\binarythreshold$ is large positive labels are only assigned to those \threads for which the initiator receives a message close to the time of the \thread publication. Thus, this threshold effectively balances whether the labeling focuses on the overall likelihood of the user receiving a message or on the likelihood of the particular post evaluated having triggered the private interaction, as we demonstrate in Section~\ref{sec:results}. 

%Recall the example from \ref{forumtimeline} in which a user creating a \thread at time $\pubtimes_{i}$ and receiving three messages after at times $\pritimes_{i,1}$, $\pritimes_{i,2}$, and $\pritimes_{i,3}$. To label this \thread, we first take all public \threads posted between $\trstart$ and $\trend$ and private messages sent to the creators of those \threads. For each private message and related \thread $\timessince$ is calculated with the precision of one minute. From this list of $\timessince$ from the set $\timesinceset$ $f_t(x)$ is derived by interpolating over the points created by taking the volume of messages per minute from $\timesinceset$. Once $f$ is derived, we evaluate it at $\pritimes_{i,1}$, $\pritimes_{i,2}$, and $\pritimes_{i,3}$. If $f(\pritimes_{i,1}) +  f(\pritimes_{i,2}) + f(\pritimes_{i,3})$ is greater than $\binarythreshold$, then the \thread is assigned a label of 1 and 0 otherwise. Again recall that $\pritimes_{i,j}$ is the difference between the time a \thread is posted and the time that the creator of that thread receives their $j^{th}$ pm after the post. 

\section{Evaluation \label{sec:eval}}

\subsection{Experimental setup}

In this section we describe the conditions in which we carry out the evaluation of our prediction method. We introduce the datasets we use and describe the preprocessing we perform prior to the experiments, as well as the machine learning tools that we use to implement the predictor.

%\paragraph{Datasets}%\ct{this paragraph first presents the three forums and characteristics. Then the duration we did on the data: only isolated posts, removed posts at the end of the leak -- they do not have responses by definition!}\ct{this text has been copied from Bekah's thesis}
\emph{Datasets.} We evaluate this method on three leaked underground forums: Carders (CC), L33tCrew (LC), and Nulled(NL). All three forums were leaked anonymously, are publicly available, and have been used in several academic~\cite{motoyama2011analysis,afrozhonor,afroz2014doppelganger,portnoff2017tools} as well as non-academic\footnote{http://krebsonsecurity.com/tag/carders-cc/} studies. We describe each forum below and we summarize their characteristics in Table \ref{tab:summary}. 

{$\bullet$ {L33tCrew}} is a German language \textit{carding} forum specializing in trading stolen credit and debit cards.  It was started in May 2007 and was leaked and closed in Nov 2009. At the time of the leak, L33tCrew had 18,834 total members, 7,687 of them participated in private message interaction. After the leak many members of L33tCrew joined Carders. 

{$\bullet$ {Carders}} is a similar German language \textit{carding} forum.  Carders was established in February 2009 and was leaked and closed in December 2010\footnote{Details of carders leak: http://www.exploit-db.com/papers/15823/}.  At the time of the leak, Carders had 8,425 total members among which 4,290 members participated in private message interaction.

{$\bullet$ {Nulled}} is a large English language forum covering a large variety of topics and is currently still active. At the time of the leak Nulled had 599,085 members. However, only 6.11\% (36,606) of the users in the leak sent or received private messages.

\begin{table}[h!]
\centering
\begin{tabular}{c c c r r r } \hline
\textbf{Forum} & \textbf{Language} & \textbf{Dates} & \textbf{Users} & \textbf{Users w/ PMs}\\ \hline
%BW  & English & 08/05-03/08 & 8718 & 1690 & 43 \\ %\hline (19.38\%) 
LC & German & 05/07-11/09  & 18834& 7687  \\ %hline (40.81\%)
CC & German & 02/09-12/10& 8425 & 4290 \\ %\hline  (50.92\%)
NL & English &01/15-05/16  & 599085 & 36606  \\  %(6.11\%)
\hline
\end{tabular}
\caption{Forums leaked data}
\label{tab:summary}
\end{table}

\begin{figure*}[h]
    \centering
    \begin{subfigure}{0.3\linewidth}%{.3\textwidth}
        \centering
        \includegraphics[width=\linewidth]{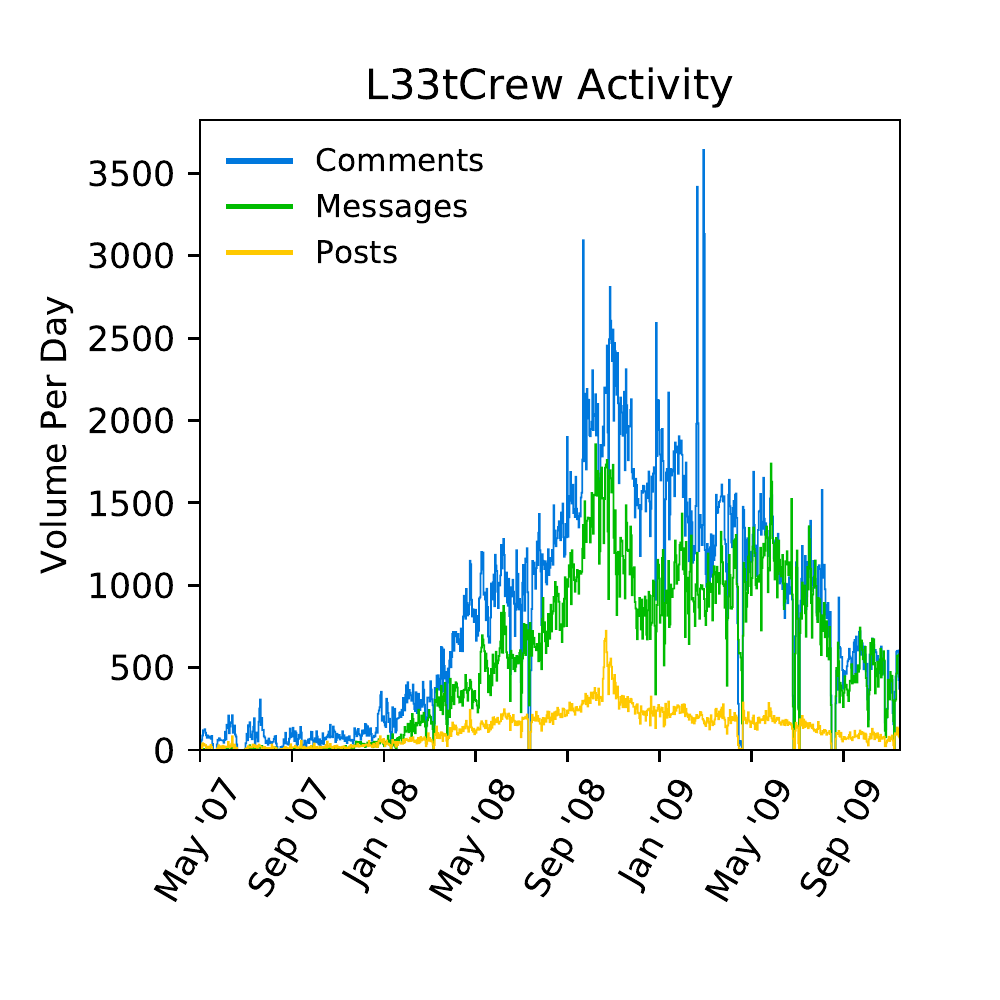}
        %\caption{L33tCrew}
    \end{subfigure}%
    \begin{subfigure}{0.3\linewidth}%{.3\textwidth}
        \centering
        \includegraphics[width=\linewidth]{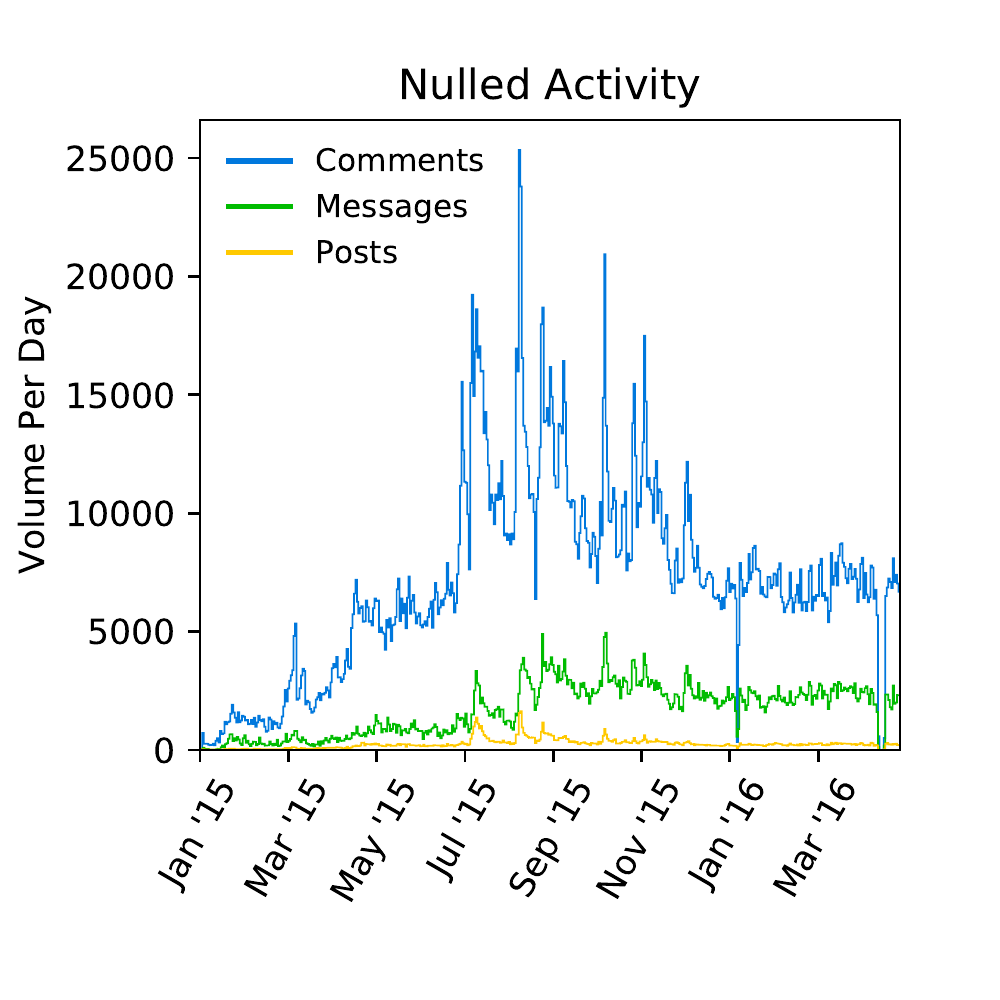}
        %\caption{Nulled}
    \end{subfigure}%
    \begin{subfigure}{0.3\linewidth}%{.3\textwidth}
       \centering
        \includegraphics[width=\linewidth]{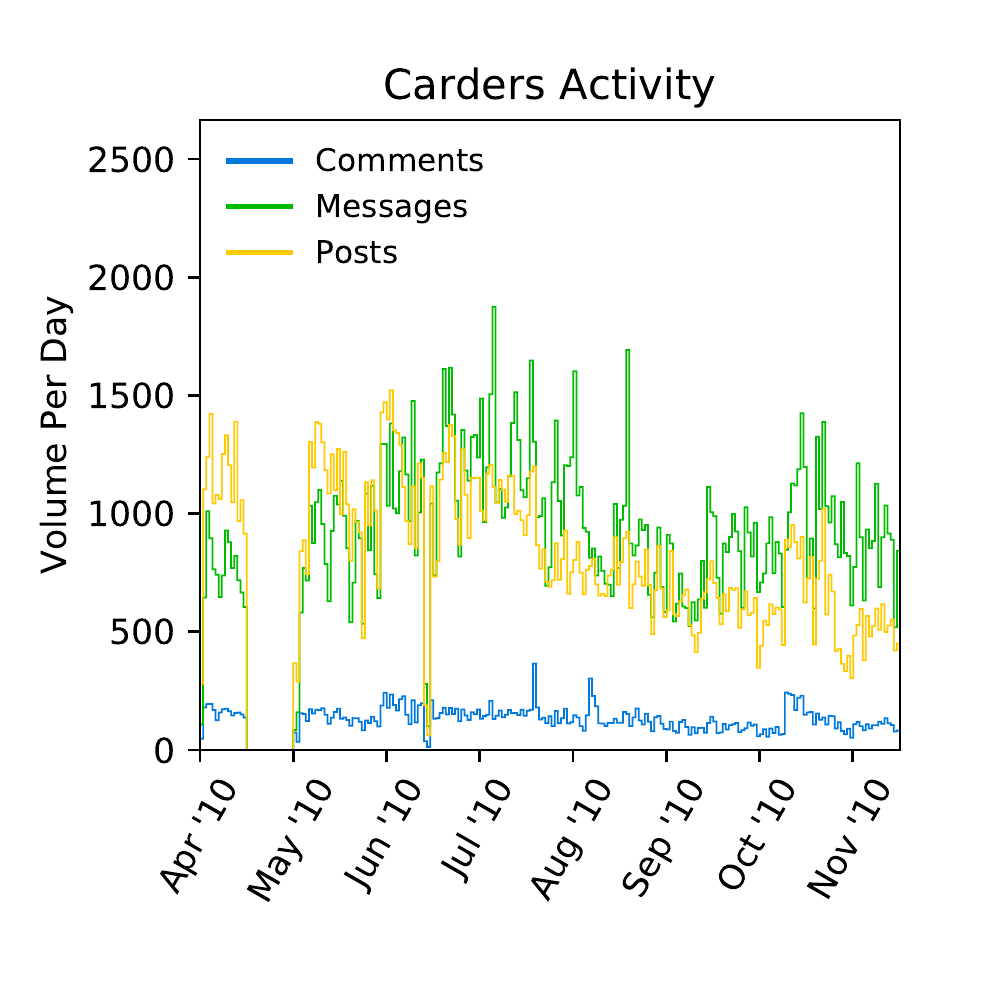}
        %\caption{Carders}
    \end{subfigure}%
    \caption{Type of activity over the time of the leaks for all forums.\label{fig:activity}}
\end{figure*}

Though all of these forums are generally similar in content and structure, the way that users interact with each of them varies. In every forum, users can create public threads, submit a comment to an existing thread, or send a private message to another user. Figure \ref{fig:activity} shows the distribution of these interactions over time for all forums. Nulled has the largest activity volume as well as a higher percentage of comments compared the posts and messages, while L33tCrew has a relatively higher proportion of private messages. Carders has much less activity than the other two, but a high volume of private interactions.

\emph{Data Pre-Processing.} 
First we select posts that are created in isolation of other posts by the same user. That is, if $u_i$ creates two posts within 12 hours of each other both are discarded. This ensures that, in training we are properly labeling the replies that correspond to that thread, and in testing we can correctly evaluate \thread classification inequivocally. Since the \thread publication times are known to the analyst this is a plausible assumption to make. 
For training we also remove \threads that are within 24 hours of the end of $\trspan$. The rationale is that most private replies to these \threads may come after $\trspan$ and are therefore unknown by definition causing the \thread to be mislabeled increasing the false negative rate. After the \thread filtering step, we select chunks of varying length $\trspan$ to simulate forum leaks that we use for training, as well as non-overlapping chunks of fixed length $\tespan$ that we use for testing, as discussed below.

%For the sake of this work, we break up each of the actual leaks into smaller chunks of length $\trspan$ to simulate training on a leak of length $\trspan$ and testing on ``unlabeled" posts. \bo{I put this here bc idk where else it should go:} In addition, we only examine posts that are created in isolation of other posts by the same user. That is, if $u_i$ creates two posts within 12 hours of each other both are discarded. We made this decision in training to ensure that we are properly labeling the replies that correspond to that thread. In the testing phase we also filter these messages out, as a clear answer to whether the posts are evaluated correctly is difficult to ascertain. Since the public messages are known to the evaluator this is a reasonable and plausible assumption to make. Finally, we remove posts that are within 24 hours of the end of $\trspan$. These training posts are likely to be mislabeled by our automated labeling method as false negatives as any associated private replies may come after $\trspan$ and are therefore unknown.  

\emph{Features \& Classifier.} 
We show in Table \ref{forumfeats} the features we use to create the models $\model$ we use for classification. We include both natural language and \thread context features. The \thread context include timing, tags, public replies tags, as well as user-related features. Among the latter we include centrality metrics from the public interaction graph, i.e., the graph that links \thread creators to \thread respondents. These centrality features have been shown in previous work~\cite{garg2015computer} to be a measure of users' popularity and influence on these forums. We run experiments with just the natural language features, just the context features, and all of the features. Note that for the natural language features we remove function words and use the frequency of the stems of the remaining words.

\begin{table}[h]
\centering
\begin{tabular}{c p{0.64\columnwidth}	}\hline
 \textbf{Feature Type}& \textbf{Description} \\ \hline

Natural Language & title bag of words, thread bag of words, and sub\-forum name bag of words \\

Context & tagged sell \thread, tagged buy \thread, time, time on forum, reply count, user reputation, views, and graph centrality metrics for the \thread creator: clustering, degree, eigenvector, betweenness \\ \hline

\end{tabular}
\caption{Summary of the feature set used for classification.}
\label{forumfeats}
\end{table}

We train a Random Forest Classifier on each of the feature groups. We choose random forest for classification for three primary reasons. First, outliers are common in our data set, and are common in data of this type generally. Second, the results are easily interpreted. That is, we can easily discover which features the decisions are made on. Finally, random forest classifiers are not as susceptible to overfitting as other similar classification techniques, which is particularly important in this problem as the training data and testing data are coming from a different time slice of the forum. Throughout our experiments we vary the classification threshold of the random forest classifier in order to create ROC curves that represent the performance of our prediction.

\emph{Simulation parameters.} To explore different scenarios, in our simulations we vary several parameters from our model, see Section~\ref{sec:predicting}. First, we vary the size of the leak, $\trspan \in [1,3,5,7]$. Smaller leaks have less \threads to train the model, and also less known private messages to use to infer the labeling function $f(x)$. For simplicity and comparability we fix the span of the testing period, $\tespan$. We empirically determined that a span of six weeks contained enough data to obtain smooth ROC curves resistant to noise.
%evaluated a variety of values for $\tespan$ and manually determined that a span of six weeks created a smooth curve, resistant to random noise in the data. Not only would an evaluator of this system likely have little control over the data they would like to evaluate, but 
Additionally, fixing $\tespan$ allows us to accurately test $\tbetween$, the time between the leak and the testing periods. Otherwise, varying $\tespan$ would affect the average time between the training and the testing \threads accross experiments. 
We also vary $\tbetween$ to evaluate the effect of time passing between the leak and testing periods. In the bulk of our experiments we use $\tbetween \in [0,3,5]$ weeks.  Since understanding how these forums change over time affects performance is vital to evaluate how well our method scales, we use an additional $\tbetween \in [10,20,30,40,50]$ for L33tCrew, the forum with the longest leak. 

Finally, we also vary the value of $\Theta$ to evaluate the performance of our method when it focuses on user-only features -- $\Theta=0$, i.e., \threads receive positive labels if the user ever receives a private message after posting); and when more weight is given to $\thread$-oriented features -- high $\Theta$, i.e., \threads receive positive labels when the initiator receives private messages close in time to the thread.
%\paragraph{Parameters} %\ct{This paragraph explains the parameters used in our experiments, which dls, deltas we chose and why. Explain how we chose the testing period}

\subsection{Results\label{sec:results}}

\begin{figure*}[h]
    \centering
    \begin{subfigure}{0.3\linewidth}
        \centering
        \includegraphics[width=\linewidth]{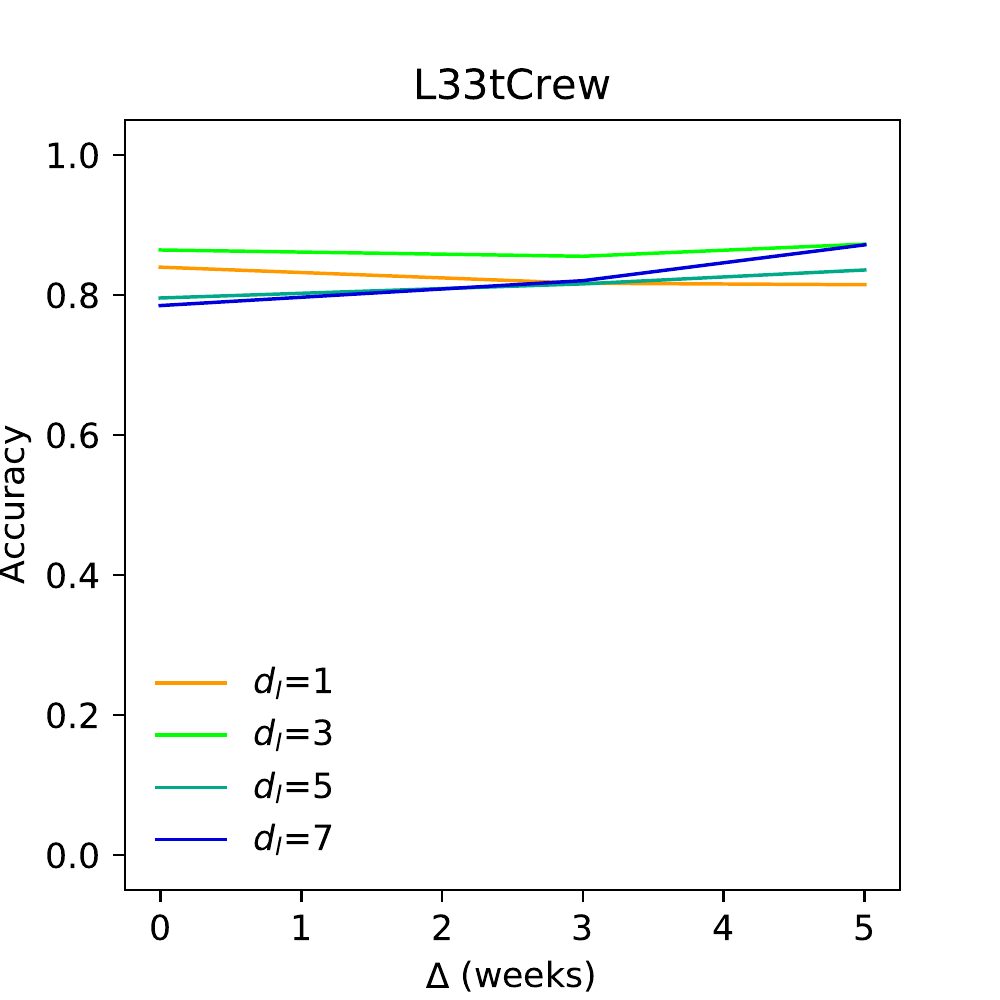}
    \end{subfigure}%
    \begin{subfigure}{0.3\linewidth}
        \centering
        \includegraphics[width=\linewidth]{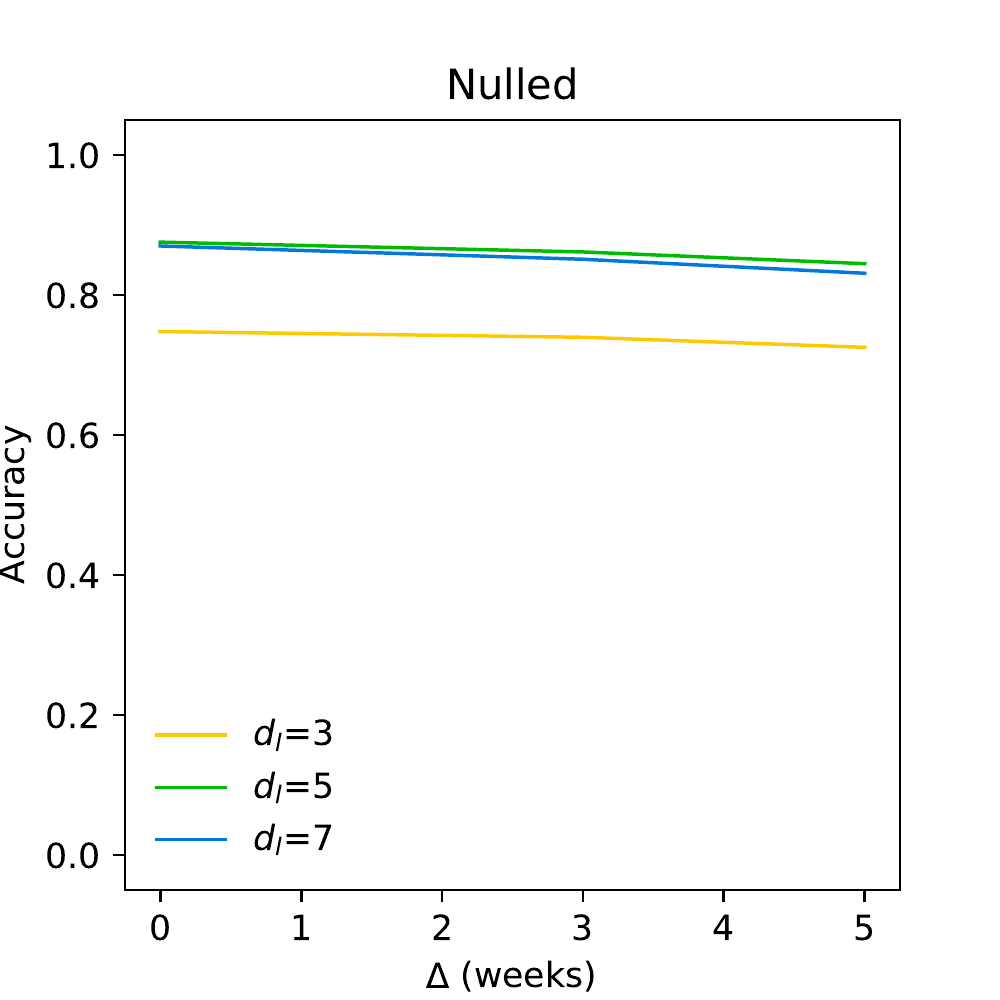}
    \end{subfigure}%
    \begin{subfigure}{0.3\linewidth}
       \centering
        \includegraphics[width=\linewidth]{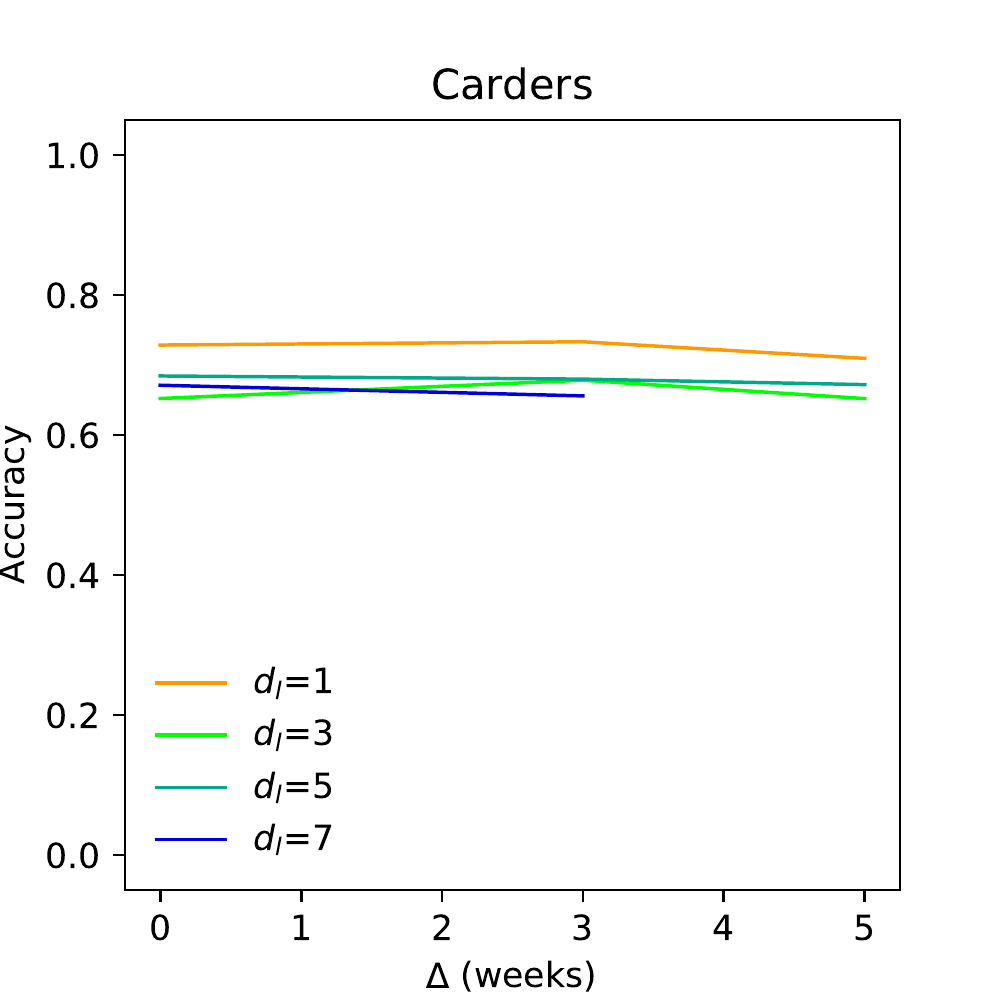}
    \end{subfigure}%
    \caption{Classification accuracy for different values of $\trspan$ and $\tbetween$ using all features. \ct{why is there no $\trspan=1$ for Nulled???} \bo{because when $d_l = 1$ week the percentage of posts for any theta, including 0, that are labeled 1 < 10\%}
\label{fig:acc}}
\end{figure*}

\paragraph{Random Forest Classifier results.} 
%\rg{clarify what theta=0 means, it is counterintuitive}
First, we study the influence of $\trspan$ and $\tbetween$ have on the accuracy of our prediction methods. We show the results in Figure~\ref{fig:acc} where for each combination of $\trspan$ and $\tbetween$ we plot the results for the largest possible $\binarythreshold$ for which the percentage of positive labels is greater than 10\%. This ensures that: i) we do not focus on user-based features that are stable over time, and ii) there is enough data in the positive class for the classification performance to be significant. We observe that varying $\trspan$, or $\tbetween$ has no notable effect on the accuracy. Even with the larger values of $\tbetween$ tested on the L33tCrew forum we see no decline in accuracy: $\tbetween=5$ and $\tbetween=40$ yielded an accuracy of 87.17\% and 81.41\%. For the rest of the experiments in this section we fix $\trspan$ to 7 weeks and $\tbetween$ to 0.

\begin{figure*}[]
    \centering
    \begin{subfigure}{0.3\linewidth}
        \centering
        \includegraphics[width=\linewidth]{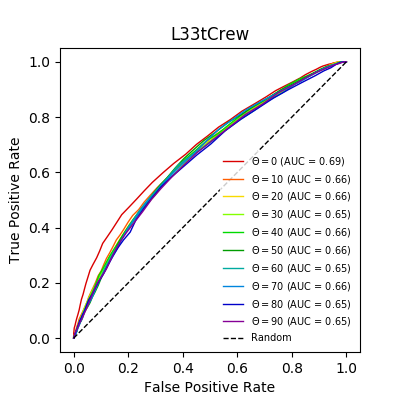}
    \end{subfigure}%
    \begin{subfigure}{0.3\linewidth}
        \centering
        \includegraphics[width=\linewidth]{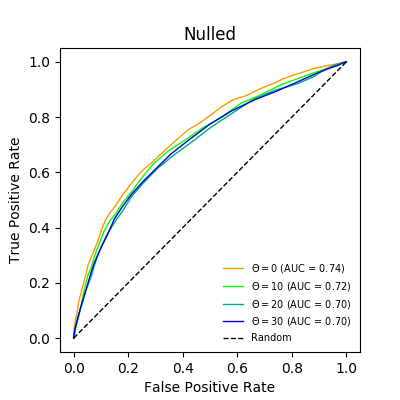}
    \end{subfigure}%
    \begin{subfigure}{0.3\linewidth}
       \centering
        \includegraphics[width=\linewidth]{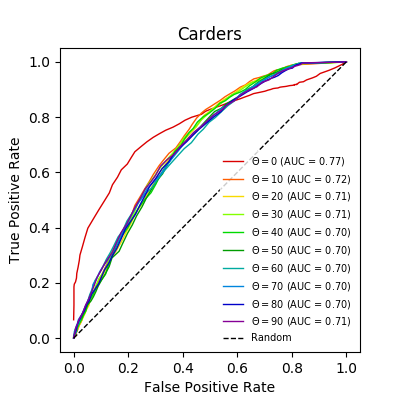}
    \end{subfigure}%
    \caption{ROC Curves for different $\binarythreshold$ values ($\trspan=7$ weeks and $\tbetween=0$).}
        \label{fig:roc}
\end{figure*}

We then study the effect of $\binarythreshold$ on performance. We plot in Figure~\ref{fig:roc} ROC curves for varying $\binarythreshold$. We see that across all three forums, and in particular for Carders, smaller values of $\binarythreshold$ where more posts are labeled as positive perform best, i.e. they result a higher True Positive Rate and a lower False Positive Rate. We conjecture that part of this success is due to small $\binarythreshold$ yielding a large percentage of positive labels, thus resulting in a more balanced training distribution. 

\begin{figure*}[]
    \centering
    \begin{subfigure}{0.3\linewidth}
        \centering
        \includegraphics[width=\linewidth]{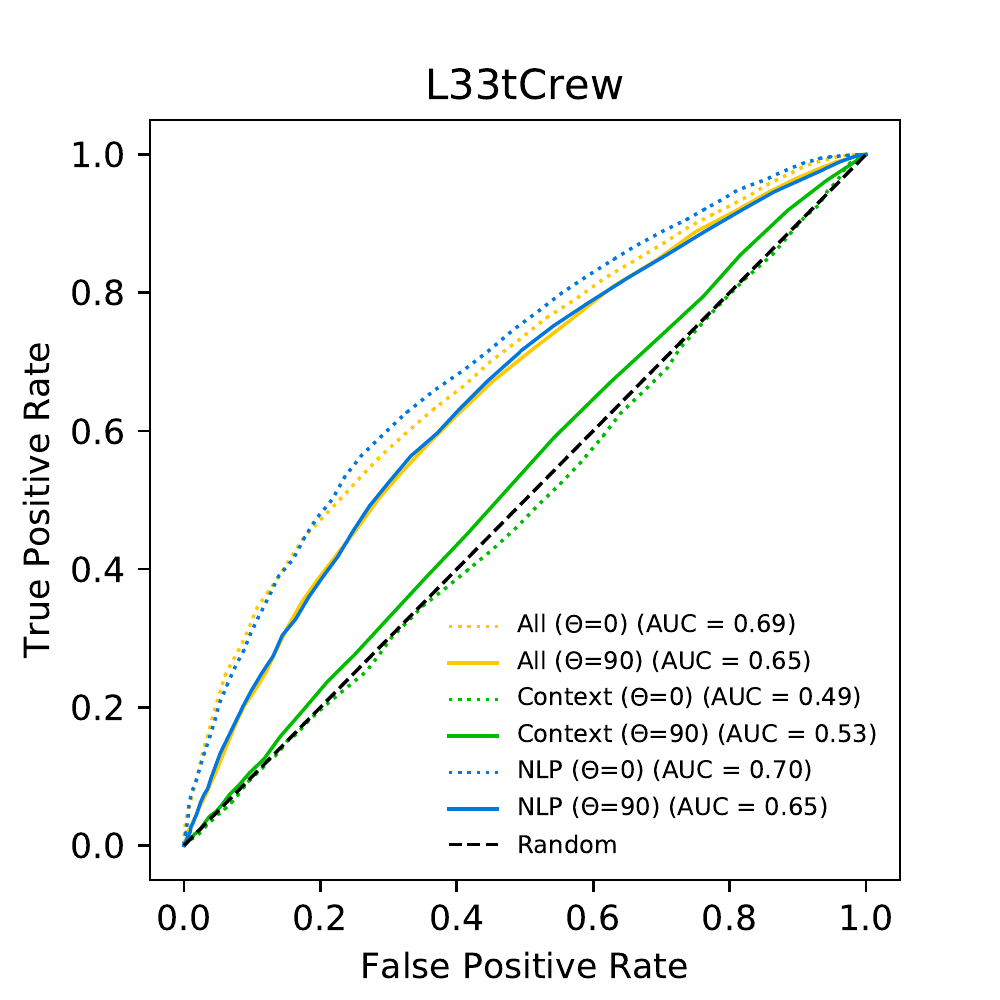}
    \end{subfigure}%
    \begin{subfigure}{0.3\linewidth}
        \centering
        \includegraphics[width=\linewidth]{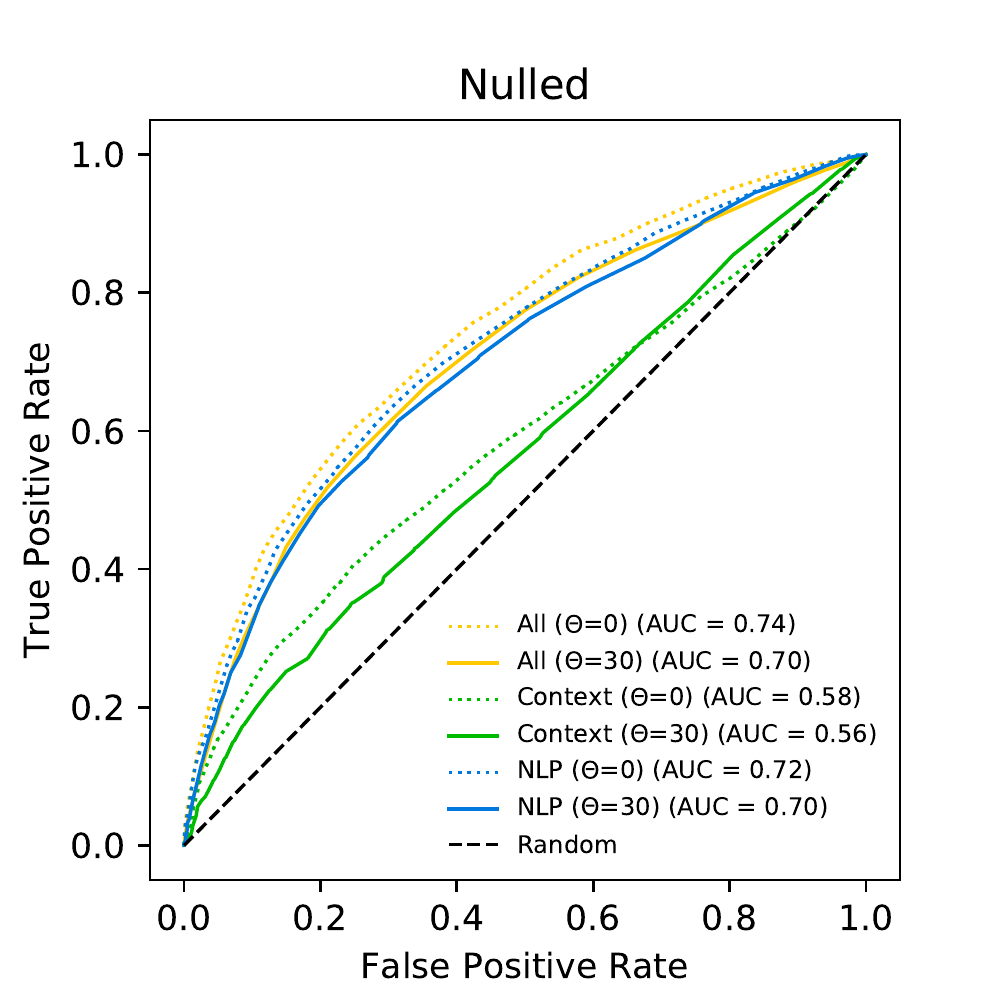}
    \end{subfigure}%
    \begin{subfigure}{0.3\linewidth}
       \centering
        \includegraphics[width=\linewidth]{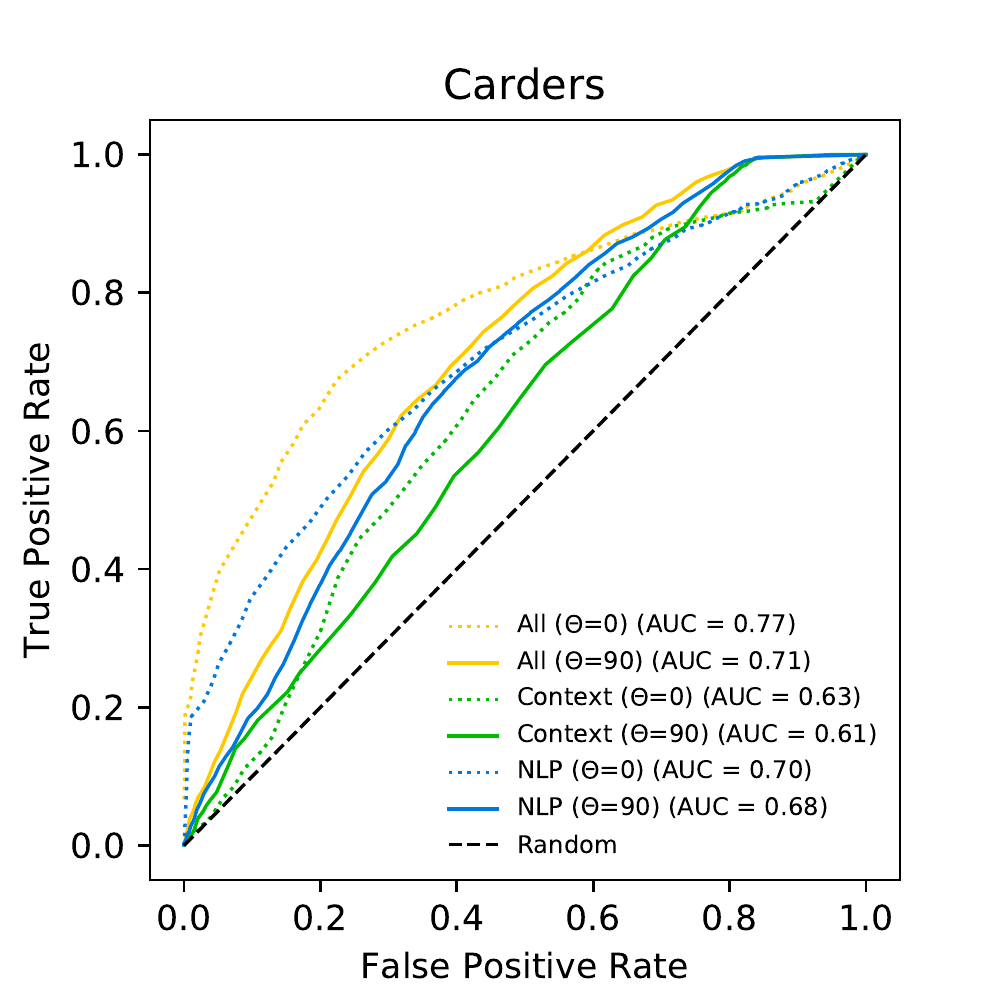}
    \end{subfigure}%
    \caption{ROC Curves for different feature sets ($\trspan=7$ weeks and $\tbetween=0$)}
        \label{fig:diff_feats}
\end{figure*}

Finally, we compare the different feature sets, see Figure~\ref{fig:diff_feats}. All of the features together perform the best, followed closely by only NLP. Both of them clearly outperform the context features alone. The context features are most useful on Carders, and on both Nulled and Carders they do better with a smaller $\binarythreshold$ than a larger one. This supports our claim that when $\binarythreshold$ is set to 0 that what is being predicted is closer to which users are more likely to receive messages rather than which \threads are likely to trigger private replies. 

\emph{Manual validation of labeling.}
We manually labeled 330 public \thread and private message pairs.  We sampled these \threads and then we selected uniformly at random one message that was sent to the \thread creator at any time in $\trspan=16$ weeks. We then labeled whether the private message appeared to be in response to the \thread. Of the 330 pairs we were able to conclusively label 175. Note that, as opposed to our automated labeling method, this manual labeling does not aim at establishing if a \thread received \emph{any} related private message, but whether a \thread created by $u_i$ and a \emph{concrete} private message sent to $u_i$ were related. Therefore, we could not expect our manual labels to fully line up with the automated labels even if both methods output perfect labels. We also label \threads in the manually labeled pairs using our automated labeler for $\binarythreshold=0$ and $\binarythreshold=90$. 

Table~\ref{tab:manual} displays the results of both labeling processes. Note that when $\binarythreshold=0$, i.e., when automated labeling assigns a positive label to each \thread creator who receives a private message in $\trspan$, there are no negatively labeled \threads for which we can label a public \thread and private message pair because there are no positively labeled \threads with private messages. That is, if we are labeling a pair with a private message to $u_i$, there is at least one message in $\trspan$, thus the label is positive by default. 

\begin{table}[h!]
\centering
\begin{tabular}{c c c c} \hline
\textbf{$\binarythreshold$} & \textbf{Automated} & \textbf{Manual}& \textbf{Percent} \\ \hline
0 & + & + & 32.93\% \\
& + & -& 67.07\% \\ \hline
%& + & ? &47.10\% \\ \hline
90 & + & + &46.15\% \\
& + & -& 53.85\% \\ \hline
 %& + & ?& 46.94\% \\ \hline
90 & - & + &17.07\% \\
 & - & - &82.93\% \\
 %& - & ? &46.75\% \\

\hline
\end{tabular}
\caption{Results of manual and automated labeling. Automated labeling configured with $\binarythreshold=0$ and $\binarythreshold=90$.}% on \threads that the automated labeler assigns positive and negative for $\binarythreshold=90$ and for which the automated labeled assigns positive for $\binarythreshold=0$.}
\label{tab:manual}
\end{table}

We see that for \threads labeled positively by the automated method with $\binarythreshold=0$ about a third of the sampled privates messages are manually labeled as related. The fact that manual labeling finds many of the posts with positive labels to have unrelated messages indicates that, indeed, the automated labeler is not focusing on the \thread itself, but to the user. Also, we note that the \threads that were manually labeled as unrelated may have a private message different from the sampled one that is actually related to the post, so the automated label may still be correct.

When we increase $\binarythreshold$ to 90, the number of posts labeled as negative increases: users that receive their messages far after the \thread will be labeled as negative, as these messages are unlikely to be related to the \thread and thus contribute little to $\likelihood{\post{i}{\tpost}}$, see \ref{eq:likelihood}, which does not reach the threshold $\binarythreshold=90$. On \threads that the automated method labeled as negative, we manually label 17\% of our pairs as related. That is, this automated labeler mislabeled these posts according to our manual labeling.

%we find that more of the \threads automatically labeled as positive have related private messages, while only few of the \threads labeled as negative have related private messages. In other words, increasing $\binarythreshold$ tunes our model to be more predictive regarding individual \threads and how likely they are to trigger private replies. 

%When $\binarythreshold=90$ more posts are labeled as negative: users that receive few messages far after the \thread will be labeled as negative, as these messages are unlikely to be related to the \thread. On \threads that the automated method labeled as negative, we only manually label 17\% of our pairs as related. That is, this automated labeler mislabeled these posts according to our manual labeling.

For the \threads that our automated method labels as positive, there is a difference between the posts labeled with $\binarythreshold=90$ and $\binarythreshold=0$. The higher percentage of pairs being manually labeled as positive for a larger $\binarythreshold$ implies that a larger $\binarythreshold$ is more likely to have a related private message than a smaller $\binarythreshold$. Thus, $\binarythreshold$ tunes the model to be more or less related to the \thread itself, or to be tuned to whether a user generally receives many messages.

\begin{figure}[h]
\centering
\includegraphics[width=\linewidth]{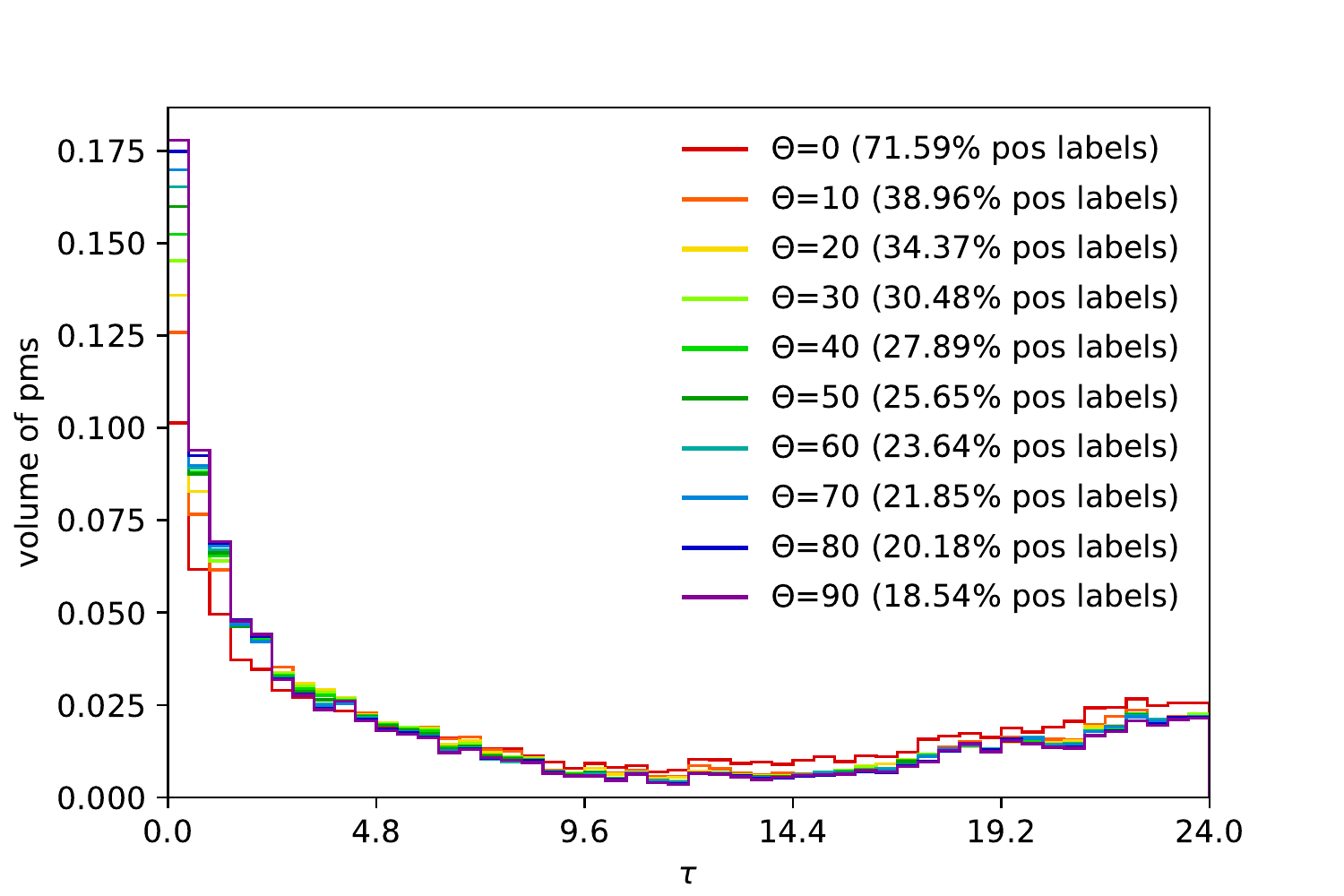}
\caption{Distributions of $\timessince$ values for Positive Labels on L33tCrew. ($\trspan=7$ weeks, $\tbetween=0$) \label{fig:taus}}
\end{figure}

Figure~\ref{fig:taus}, where we plot the distribution of $\timessince$ values for positively labeled \threads, supports this claim. We observe that for $\binarythreshold=0$ the volume of messages at small $\timessince$ is smaller than for other $\binarythreshold$, but larger for large $\timessince$. In other words, large $\binarythreshold$ considers a much larger percentage of private messages sent closer to the $\thread$ publication time, and therefore likely to be related to it. On the other hand small $\binarythreshold$ consider messages all across $\trspan$, i.e., considers only whether the message has been sent to the target receiver regardless of the time. 
%line is shorter than the rest in the beginning and taller as time passes. This trend continues with larger $\timessince$ values. Larger $\binarythreshold$'s are more sensitive to time. \bo{I'll finish this explanation later.}

\bo{put this somewhere: Note that for this analysis $\trstart$ is always the midpoint of the data available.  }\ct{Why is this important?} \bo{bc experiment setup details are important for reproducibility}

\emph{Feature Analysis.} Next we aim to determine which features are the best predictors, for which we calculate the information gain at each experiment. Surprisingly, we find that the highest ranked features in most experiments are the centrality metrics, number of public replies, number of views, and how long the user has been on the forum. We observe no notable difference between the top ranked features for different values of $\binarythreshold$. This seem to contradict the results of the Random Forest Classifier which show that performance is better using the NLP features. This discrepancy is due to the volume of these features. Though context features contain a lot of information and individually are very predictive, they are only a handful. On the other hand, we have thousands \ct{correct?} of NLP features, thus they are more predictive as an ensemble.

\begin{figure}[h]
\centering
\includegraphics[width=0.65\linewidth]{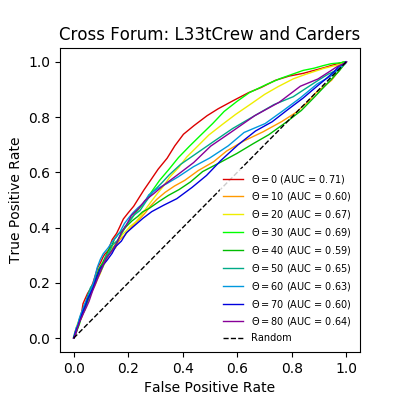}
\caption{Cross Forum Results for L33tCrew and Carders.\label{fig:crossforumres}}
\end{figure}

\emph{Cross-Forum Classification} \label{sec:ufcrossforum} Here we consider the case where the analyst wants to predict private activity on a forum for which she has no leaked data, but a similar forum exists with leaked private message data. We take as examples L33tCrew and Carders since they are the most similar forums considered in this work, both being German language carding forums. We train the classifier on data from Carders and test it on data from L33tCrew. The results are displayed in Figure \ref{fig:crossforumres}.  While the classifier does outperform random chance, the results are, predictably, worse than the intra-forum results presented above. 

\section{Discussion}
  
%In this section we discuss the limitations of our supervised machine learning based method and its envisioned use cases, along with the ethical framework of our study.

\emph{Limitations.}
The accuracy of our method is not exceptionally high, meaning that its results cannot be used to directly act upon a \thread or user. However, we have found that predicting private activity is a difficult problem and even manual analysis often cannot ascertain if a private message was sent in response to public \threads. Thus, providing analysts with automated tools to point out where to concentrate their efforts is of value. 

Another key limitation of our method is that, though it is quite effective at predicting which \threads will be followed by private interactions, it cannot predict who will participate in the private communication. %In other words, it cannot predict the graph of members that privately communicated with each other on a forums. 
Such a graph would be useful for understanding collaborations among members and their relations. However, the revelation of our analysis that there is little overlap between members whom post in the same public \thread and those whom communicate privately hints that inferring such graph is a very complex problem.

Also, our analysis has focused on `isolated' \threads, i.e., \threads for which the poster does not start any other \thread in within 12 hours. While this simplification still allows us to show that private interactions can indeed be predicted, it limits the applicability of our approach. More research is needed to develop means to jointly infer the likelihood of responses when a series of \threads are published, and to devise whether it is possible to decide which of these public communications has, or have, actually triggered the private interaction.

Finally, we observe that, although our technique outperforms random guessing when trained on one forum and tested on another forum, it currently performs better when there is training data for the target forum. %In other words, there is a variation in users' behavior between forums. 
This lays out an interesting problem to be addressed in future work, focused on how to identify stable features that can be used to bridge the difference in domains posed by the diversity between different forums' users.
%improve performance when there is no private training data for a target forum. One idea is that public data might be able to improve accuracy if public features can be identified that correlate to private ones used in the training process.

\emph{Use Cases.}
Even though our approach has several limitations there are useful and realistic use cases. The first envisioned use case is for forums such as Nulled, where it continues to operate after the forum leak. Researchers or private companies could continue collating the public messages from Nulled or another forum that continued operating after a leak. Researchers or analysts can use our approach to identify public \threads that likely generated private interactions. This could enable measures of forum member centrality based on private interaction or identify members and public \threads that should be further examined based on likely private interactions which might indicate completed sales.

%While our approach is far from a perfect predictor, we envision that an analyst could use this method with a declining $\binarythreshold$ to flag public posts for further study.

\emph{Ethics.}
All of the data used in our study was publicly leaked and believed to be authentic. Our IRB deemed our study exempt based on the fact that the data was publicly leaked and that by the nature of the data being underground criminal forums it is unlikely anyone used their real name or other Personal Identifiable Information (PII). Our analysis of private messages was focused on identifying links between private messages and \threads, and not analyzing users identities or messages content. If we had discovered PII in these messages we would have notified our IRB and submitted an amended IRB application. However, we did not find any PII during our manual analysis.

\section{Conclusions}
%\dm{this is some placeholder text, I or someone else needs to clean this up}

Private activity in underground forums has been shown to be key for understanding the underlying cybercriminal ecosystem. However, analysts rarely have access to private messages and when they do these private interactions are limited to a snapshot that only represents forum activity during a bounded time interval. 

In this work we propose and evaluate a method that provides analysts with the means to predict such private interactions from the information that is  available to them. We have presented a supervised machine learning based method able to predict which public \threads will generate private messages, after a partial leak of such messages has occurred. Additionally, we have proposed a method for automated labeling that reduces the cost of our analysis, thus increasing its potential to be deployed to analyze large forums. This automatic labeling method has a parameter %$\binarythreshold$ that can be tuned to focus the model more on a user's likelihood of generating responses (low $\binarythreshold$) or on a post's likelihood of generating responses (high $\binarythreshold$). 
This is especially useful as manual labeling turned out to be even more difficult than expected.

To the best of our knowledge, our approach is the first proposed to tackle this important underground forum analysis problem of limited to no information about private interactions in most underground forums. Our work is an initial step to provide forum analysts with tools to understand which posts are likely result in sales or follow up discussion and are more likely important for them to spend more time investigating. Understanding how this learning problem transfers across forums is a crucial part of future work in this area.

\bibliographystyle{ACM-Reference-Format}
\bibliography{ref} 

\end{document}